\documentclass[useAMS,usenatbib]{mn2e}
\usepackage{graphicx}

\newcommand       \be           {\begin{equation}}
\newcommand       \ee           {\end{equation}}
\newcommand       \bea          {\begin{eqnarray}}
\newcommand       \eea          {\end{eqnarray}}
\newcommand       \apj          {ApJ}
\newcommand       \apjs         {ApJS}
\newcommand       \araa         {Annual Reviews of Astronomy \& Astrophysics}
\newcommand       \aj           {AJ}
\newcommand       \apjl         {ApJL}
\newcommand       \aap          {A\&A}
\newcommand       \nat          {Nature}
\newcommand       \mnras        {MNRAS}
\def\simlt{\mathrel{\hbox{\rlap{\hbox{\lower4pt\hbox{$\sim$}}}\hbox{$<$}}}}
\def\simgt{\mathrel{\hbox{\rlap{\hbox{\lower4pt\hbox{$\sim$}}}\hbox{$>$}}}}
\def\simlt{\mathrel{\hbox{\rlap{\hbox{\lower4pt\hbox{$\sim$}}}\hbox{$<$}}}}
\def\simgt{\mathrel{\hbox{\rlap{\hbox{\lower4pt\hbox{$\sim$}}}\hbox{$>$}}}}
\def\simlt{\mathrel{\hbox{\rlap{\hbox{\lower4pt\hbox{$\sim$}}}\hbox{$<$}}}}
\def\simgt{\mathrel{\hbox{\rlap{\hbox{\lower4pt\hbox{$\sim$}}}\hbox{$>$}}}}
\def\lesssim{\mathrel{\hbox{\rlap{\hbox{\lower4pt\hbox{$\sim$}}}\hbox{$<$}}}}
\def\gtrsim{\mathrel{\hbox{\rlap{\hbox{\lower4pt\hbox{$\sim$}}}\hbox{$>$}}}}

\title[]{Relic proto-stellar disks and the origin of luminous circumstellar interaction in core collapse supernovae} \author[B.~D. Metzger]{B.~D. Metzger$^{1,2}$ \thanks{E-mail: bmetzger@astro.princeton.edu}\\ $^{1}$Department of Astrophysical Sciences, Peyton Hall, Princeton University; Princeton, NJ 08544 USA \\ $^{2}$NASA Einstein Fellow}
\begin{document}
\date{Accepted . Received ; in original form }
\pagerange{\pageref{firstpage}--\pageref{lastpage}} \pubyear{????}
\maketitle
\label{firstpage}

\begin{abstract}
A small fraction of core collapse supernovae (SNe) show evidence that the outgoing blast wave has encountered a substantial mass $\sim 1-10M_{\sun}$ of circumstellar matter (CSM) at radii $\sim 10^{2}-10^{3}$ AU, much more than can nominally be explained by pre-explosion stellar winds.  In extreme cases this interaction may power the most luminous, optically-energetic SNe yet discovered.  Interpretations for the origin of the CSM have thus far centered on explosive eruptions from the star just $\sim$ years$-$decades prior to the core collapse.  Here we consider an alternative possibility that the inferred CSM is a relic disk left over from stellar birth.  We investigate this hypothesis by calculating the evolution of proto-stellar disks around massive stars following their early embedded phase using a self-similar accretion model.  We identify a brief initial gravitationally-unstable (``gravito-turbulent'') phase, followed by a much longer period of irradiation-supported accretion during which less effective non-gravitational forms of angular momentum transport dominate.  Although external influences, such as the presence of a wide binary companion, may preclude disk survival in many systems, we find that massive ($\sim 1-10M_{\sun}$) disks can {\it preferentially survive around the most massive stars}.  Reasons for this perhaps counter-intuitive result include (1) the shorter stellar lifetimes and (2) large photo-evaporation radii ($\sim 10^{3}$ AU) of very massive stars; (3) suppression of the magneto-rotational instability due to the shielding from external sources of ionization; and (4) relative invulnerability of massive disks to lower mass stellar collisions and luminous blue variable eruptions.  Furthermore, disks with radii $\sim 10^{2}-10^{3}$ AU are picked out by the physics of the embedded stage of accretion and the requisite conditions for subsequent disk survival.  The disk mass, radius, and scale-height at core collapse typically result in a $\sim 10$ per cent efficiency for converting the kinetic energy of the exploding star into radiation, potentially producing a total electromagnetic output $\sim 10^{50}-10^{51}$ ergs.  We identify two regimes of disk-supernova interaction, which are distinguished by whether the shocked disk radiates its thermal energy before being engulfed by the expanding SN ejecta.  This dichotomy may explain the difference between very luminous supernova which show narrow H line emission and those which show no direct evidence for hydrogen-rich CSM interaction.  Because very luminous SNe are rare, testing the relic disk model requires constraining the presence of long-lived disks around a small fraction of very massive stars.

\end{abstract}

\begin{keywords}
{stars: winds, outflows, formation; circumstellar matter; supernova: general; accretion: accretion disks; infrared: stars}
\end{keywords}

\section{Introduction}

Sensitive, wide-field transient surveys are revolutionizing our understanding of the breadth of time-dependent astrophysical phenomena.  The discovery of extremely sub-luminous, rapidly-evolving Type I supernovae (SNe) (e.g. \citealt{Foley+09}; \citealt{Valenti+09}; \citealt{Perets+09}; \citealt{Poznanski+10}) may, for instance, indicate a wider diversity of white dwarf-related explosions than previously anticipated (\citealt{Bildsten+07}; \citealt{Metzger+09}; \citealt{Shen+10}).  In the opposite extreme, a number of core-collapse SNe have recently been discovered that are unusually bright and/or optically-energetic.  These events, collectively known as very luminous SNe (VLSNe), include SN 2006gy (\citealt{Ofek+07}; \citealt{Smith+07}), 2006tf (\citealt{Smith+08}), 2005ap (\citealt{Quimby+07}), 2003ma (\citealt{Rest+09}), 2007bi (\citealt{Gal-Yam+09}), 2008fz (\citealt{Drake+09}), 2008es (\citealt{Gezari+09}; \citealt{Miller+09}) and 2008iy (\citealt{Miller+10}).  The mysterious event SCP 06F6 (\citealt{Barbary+09}) and the luminous, high-redshift transients discovered by the Palomar Transient Factory \citep{Quimby+09} may be related phenomena.  VLSNe radiate a total energy $\gtrsim 10^{50}-10^{51}$ ergs and can reach peak absolute magnitudes approaching $M_{\rm V} \sim -23$ (e.g. 2005ap).  Though they represent only a small fraction of massive stellar deaths, VLSNe are thus particularly conspicuous because they can remain bright for months$-$years and are detectable even in the distant universe (\citealt{Cooke+09}; \citealt{Quimby+09}).  

The optical emission from most Type I and some Type II SNe is powered by the radioactive decay sequence $^{56}$Ni $\rightarrow$ $^{56}$Co $\rightarrow$ $^{56}$Fe, suggesting that VLSNe may result from an anomalously high Ni yield (e.g. at least several solar masses are required to explain SN 2006gy; \citealt{Kawabata+09}).  Though such a large $^{56}$Ni mass  is difficult to produce in a normal core collapse explosion, it is a natural byproduct of theoretically predicted ``pair-instability'' SNe (PI-SNe; \citealt{Barkat+67}; \citealt{Bond+84}).  Although PI-SNe are a plausible origin for some events including SN 2006gy \citep{Smith+07} and SN 2007bi \citep{Gal-Yam+09}, they are an unlikely explanation for all VLSNe since PI-SNe are expected to occur only from extremely massive stars with low metallicities \citep{Heger&Woosley02}, as would form primarily in the early universe.

Another way to power an anomalously bright SN light curve is to increase the explosion's prompt radiative efficiency; indeed, the optical output of most core collapse SNe ($\sim 10^{49}$ ergs; e.g. \citealt{Bersten&Hamuy09}) represents only a small fraction of the total $\gtrsim 10^{51}$ ergs of kinetic energy that is generally available.  Low radiative efficiency is typical because the stellar progenitors of core-collapse SNe are relatively compact: radii for Wolf-Rayet and red giants are $R \sim 0.01$ and $\lesssim 10$ AU, respectively.  Much of the thermal energy generated by the shocked stellar envelope is therefore lost to PdV work before the ejecta expands sufficiently to become transparent.

The radiative efficiency can be enhanced, however, if the kinetic energy of the ejecta is thermalized at much larger radii ($\gtrsim 100$ AU), in a region above or closer to the photosphere (e.g. \citealt{Falk&Arnett73,Falk&Arnett77}; \citealt{Chevalier&Fransson94}; \citealt{Smith&McCray07}).  This can occur via shock interaction with a dense circumstellar medium (CSM).  Indeed, the Type IIn class of SNe (e.g. \citealt{Schlegel90}; \citealt{Filippenko97}), which are characterized in part by narrow H emission lines suggestive of slower material, are generally believed to be powered by CSM interaction (e.g. \citealt{Chugai+04}).  Although many VLSNe are indeed classified as Type IIn, others show little or no evidence for CSM interaction (such as SN 2005ap; \citealt{Quimby+07}) or hydrogen in their spectra.  Such behavior may nevertheless be consistent with a CSM-powered luminosity if the CSM is especially massive and opaque, since this would delay the time required for the shocked thermal energy to escape \citep{Smith&McCray07} and for the SN to begin displaying more typical IIn properties (e.g. \citealt{Smith+08}).  We describe a specific example of this effect in $\S\ref{sec:tworegimes}$.  

In order to explain luminous IIn SNe ranging from events like 2008iy \citep{Miller+10} to extreme cases such as SN 2006gy via CSM interaction requires the presence of $\sim 0.1-10M_{\sun}$ on radial scales $\sim 100-1000$ AU.  If one assumes that the narrow H line widths ($\sim 10^{2}-10^{3}$ km s$^{-1}$) commonly observed in SN IIn represent the unshocked outflow speed of the {\it same} CSM responsible for powering the SN luminosity, the inferred mass-loss rates range from $\sim 10^{-3}M_{\sun}$ yr$^{-1}$ up to $\sim M_{\sun}$ yr$^{-1}$, much higher than can be supplied by standard, line-driven winds (e.g. \citealt{Gayley+95}; \citealt{Owocki+04}; \citealt{Smith&Owocki06}).  When combined with evidence for minima or ``gaps'' in the mass distribution between the star and the CSM in some cases (e.g. \citealt{Chugai&Danziger94}; \citealt{Miller+10}), these inferences have motivated the interpretation that the CSM is composed of shells of material ejected just years$-$decades before the SN explosion (e.g. \citealt{Chugai+04}; \citealt{Smith+07}; \citealt{Dessart+09}).    

Proposed explanations for these pre-SN outbursts thus far include luminous blue variable (LBV)-like eruptions analogous to the 19th century eruption of $\eta$-Carinae (e.g.~\citealt{Smith+07}) and pulsational pair instabilities (i.e. sub-energetic, non-terminal PI-SN analogs; \citealt{Woosley+07}).  However, neither of these explanations is altogether satisfactory.  If LBV eruptions do occur just prior to stellar death, this suggests that very massive stars can die in their LBV phase, nominally thought to occur prior to core He burning (e.g. \citealt{Langer+94}), and with their hydrogen envelopes intact (e.g. \citealt{Smith08a}).  Pair-instabilities, on the other hand, occur only in stars with masses $\gtrsim 95 M_{\sun}$ that retain their He envelope following H burning; this again likely requires low metallicity, an unusual circumstance in the present-day universe.  

It has been argued that both the hydrogen-rich deaths of very massive stars and modern-day PI-SNe become more likely if empirically-calibrated stellar mass-loss rates have been systematically overestimated (\citealt{Gal-Yam+07}; \citealt{Smith08b}) due to the insensitivity of $n^{2}$ emission diagnostics to wind clumping (e.g.~\citealt{Fullerton+06}).  This does not, however, mitigate the fact that both LBV and pair instability models also appear to require fine-tuning: the delay between shell ejection and core collapse must be well-timed to produce a collision at the optimal radii ($\sim 10^{2}-10^{3}$ AU) for producing bright emission.  Although this ``coincidence'' could in part result from a selection effect, the fact remains that the pre-SN eruption and core collapse {\it must be preferentially synchronized to within a timescale $\sim$ years$-$decades (which is orders of magnitude shorter than stellar transport or evolutionary timescales) or VLSNe would be even more rare than is observed}.  Although such a correlation could in principle arise from the time interval between pair-instabilities, this depends sensitively on uncertain details such as how long the core takes to contract following the first pulse (\citealt{Heger&Woosley02}; \citealt{Woosley+07}).

In this paper we consider an alternative origin for CSM interaction in luminous core collapse SNe: a gaseous proto-stellar disk left over from stellar birth.  The lifetimes of the most massive stars are only $\sim 3$ Myr (e.g. \citealt{Bond+84}; \citealt{Maeder&Meynet87}), comparable to the observed lifetimes of disks around low-mass proto-stars (e.g. \citealt{Strom95}).  Observations of moderately massive ($\sim 10 M_{\sun}$) proto-stars typically suggest much shorter disk lifetimes (e.g. \citealt{Natta+00}; \citealt{Fuente+06}; see, however, \citealt{Manoj+07}), indicating that the ``disk dispersal'' processes at work following star formation may be more effective for higher mass stars (e.g. \citealt{Hollenbach+00}).  

VLSNe are, however, very rare and likely originate from a small subset of very massive stars.  The PTF events discovered by \citet{Quimby+09} represent only a tiny fraction $\sim 10^{-4}$ of core collapse SNe.  Although Type IIn SNe as a whole account for a few to $\sim 10$ per cent of core collapse events (\citealt{Cappellaro+99}), they are a highly inhomogeneous population and the majority may represent a different phenomena than the most luminous events.  If a fraction $f \sim 10^{-4}-10^{-2}$ of massive stars must retain a massive disk in order to be consistent with the rate of VLSNe, then only a few to several hundred relic disk systems should be present in Milky Way-type galaxies at any time.  Testing this prediction is nontrivial because although a relic disk would likely be conspicuous, massive stars are necessarily very young and, hence, typically far away and often obscured; observations and their interpretation are thus particularly challenging due to limited sensitivity, angular, and spectral resolution (e.g. \citealt{Cesaroni+07}).  Although there is no evidence for long-lived massive disks around unobscured ZAMS O stars, relic disk systems may not appear as normal or unobscured, and the local census of massive obscured stars and their environments is incomplete (e.g.~\citealt{Wachter+10}).  Indeed, the rate of VLSNe is sufficiently low that if only a very small fraction of high mass stars retain a disk, it would have important observable implications.

The present theoretical study can broadly be divided into two parts: a detailed analysis of (1) the conditions under which a massive proto-stellar disk can survive until the core collapse of its host star; (2) the interaction of the relic disk (with properties determined self-consistently) with the outgoing supernova ejecta and the resulting emission.  In detail, the following sections are organized as follows.  In $\S\ref{sec:stages}$ we discuss the stages in the formation and evolution of massive proto-stellar accretion disks.  Relying heavily on the results from $\S\ref{sec:stages}$, in $\S\ref{sec:models}$ we construct time-dependent proto-stellar disk models which focus on the isolated stages of evolution following the embedded accretion phase, neglecting for the moment all sources of disk mass loss except accretion.  In $\S\ref{sec:destruct}$ we examine the susceptibility of the disk to destructive processes such as outflows and stellar collisions in order to determine whether and under what conditions disk survival is most probable.  In $\S\ref{sec:diskshock}$ we explore the interaction of the disk with the supernova ejecta following core collapse and its implications for the origin of luminous CSM-powered SNe.  In $\S\ref{sec:discussion}$ we discuss the implications of our results for long-lived disks around massive stars ($\S\ref{sec:disks}$), Type IIn (and {\it non-}IIn) VLSNe ($\S\ref{sec:type2n}$) and ``hybrid'' Type I/IIn SNe ($\S\ref{sec:hybrid}$).  We conclude in $\S\ref{sec:conclusions}$.  

\section{Stages of Disk Evolution}
\label{sec:stages}
This section provides a discussion of the formation and evolution of proto-stellar disks around very massive stars.  We begin in $\S\ref{sec:embed}$ by describing the embedded phase, during which the proto-star grows by active accretion from its parent molecular core.  During this stage, most of the disk is gravitationally unstable; the outer disk is susceptible to fragmention and the inner disk rapidly feeds gas (and, possibly, bound companions) onto the central proto-star.  Once infall shuts off, the embedded phase ends and the remaining disk viscously evolves in relative isolation (neglecting for the moment external destructive processes; $\S\ref{sec:destruct}$).  This isolated evolution generally begins, as during the disk's embedded evolution, with a gravitationally-unstable phase, during which ``{\it gravito-turbulence}'' supports the disk against fragmentation and supplies the angular momentum transport ($\S\ref{sec:GIphase}$).  This phase is short-lived, however, because as accretion depletes the mass of the disk, stellar irradiation becomes an increasingly important source of midplane pressure.  The disk thus rapidly transitions into a gravitationally stable, {\it irradiation-supported} accretion phase ($\S\ref{sec:irrad}$).  From this point on, other generally less efficient processes, such as MHD turbulence under partially-ionized conditions, supply the angular momentum transport.  This typically results in a much slower disk evolution until stellar core collapse.  

\subsection{Disk Formation and the Embedded Phase}
\label{sec:embed}

The formation of a centrifugally-supported disk may be a nearly ubiquitous feature of stellar birth due to the substantial angular momentum of the parent molecular gas.  For low mass proto-stars such as T-Tauri stars, evidence for disks is strong and sometimes explicit (e.g. \citealt{Bertout89}; \citealt{Burrows+96}; \citealt{Simon+00}).  Although observations are generally more challenging for high mass stars (e.g. \citealt{Beuther+06}; \citealt{Cesaroni+07}), advances in submillimetre observations of massive protostars ($\gtrsim 10M_{\sun}$) have revealed the presence of flattened structures and, possibly, Keplerian disks (e.g. \citealt{Chini+04}; \citealt{Cesaroni+05}; \citealt{Patel+05}).  In fact, disk accretion is probably crucial to the very process by which very massive stars form: a disk shields infalling material and re-directs the stellar radiation field (e.g. \citealt{Stahler+00}; \citealt{Yorke+02}; \citealt{Krumholz+05}), allowing accretion despite the stifling effects of radiation pressure on gas and dust (e.g. \citealt{Kahn74}; \citealt{Wolfire&Cassinelli87}).

During the earliest, ``embedded'' phase in massive star formation, the protostar accretes most of its final stellar mass $M_{\star} \sim 10-100M_{\sun}$ from its progenitor molecular core on a timescale $t_{\rm acc} \sim 1-2\times 10^{5}$ years, corresponding to an accretion rate $\dot{M}_{\rm c} \sim M_{\star}/t_{\rm acc} \sim 10^{-4}-10^{-3}M_{\sun}$ yr$^{-1}$ (e.g. \citealt{McKee&Tan03}; \citealt{Banerjee&Pudritz07}).  Such a large accretion rate cannot in general be accommodated by local viscous torques due to e.g. hydrodynamical or MHD turbulence \citep{Balbus&Hawley98}.  However, once the mass of the disk grows to a substantial fraction of the mass of the central proto-star, the disk becomes susceptible to gravitational instabilities (GI; e.g. \citealt{Tomley+91}; \citealt{Johnson&Gammie03}).  These generally set in once the \citet{Toomre64} parameter 
\be
Q = \frac{c_{s}\Omega}{\pi G \Sigma} \approx \frac{M_{\star}}{\pi R^{2}\Sigma}\frac{H}{R}
\label{eq:ToomreQ}
\ee
decreases below a critical value $Q_{\rm 0} \sim 1$.  Here $c_{\rm s}$ is the adiabatic sound speed, $\Omega = (GM_{\star}/R^{3})^{1/2}$ is the orbital frequency (assuming a Keplerian potential), $\Sigma$ is the surface density, $M_{\star}$ is the stellar mass, $\pi R^{2}\Sigma$ is a measure of the local disk mass at radius $R$, and $H = c_{\rm s}/\Omega$ is the disk scaleheight, assuming vertical hydrostatic balance.

Numerical simulations show that the nonlinear development of GI depends on the thermodynamic properties of the disk (e.g. \citealt{Gammie01}; \citealt{Lodato&Rice04}).  If the disk cooling time $t_{\rm cool} \sim \Sigma c_{\rm s}^{2}/F$ exceeds the orbital period $\Omega^{-1}$, where $F$ is the disk's outward vertical energy flux, the disk settles into a quasi-steady state of ``gravito-turbulence.'' Dissipation of this GI-induced turbulence heats the disk (thus raising $c_{\rm s}$) until $Q \sim Q_{0}$, i.e. marginal gravitational stability obtains (e.g. \citealt{Gammie01}; \citealt{Rafikov05,Rafikov07,Rafikov09}; \citealt{Matzner&Levin05}; \citealt{Boley+06}).  The effective dimensionless \citet{Shakura&Sunyaev73} ``alpha'' parameter corresponding to the requisite level of turbulent viscosity was found by \citet{Gammie01} to be
\be
\alpha_{\rm gi} = \frac{4}{9(\gamma-1)\Omega t_{\rm cool}},
\label{eq:alphaGI}
\ee
where $\gamma$ is the adiabatic index.

When, on the other hand, $t_{\rm cool} \lesssim \Omega^{-1}$, gravitational instabilities cause the disk to fragment into bound substructures.  One way to interpret this result from equation (\ref{eq:alphaGI}) is that the disk cannot provide stress at the level $\alpha \gtrsim 1$ (e.g.~\citealt{Rice+05}).  The critical conditions for fragmentation, viz.~$\alpha_{\rm gi} \gtrsim 1$ for $Q = Q_{0} \sim 1$, can be translated into a maximum accretion rate $\dot{M}_{\rm max} \approx 3\pi\nu\Sigma \sim c_{\rm s}^{3}/G$ that the disk can accommodate without fragmenting, where $\nu = \alpha_{\rm gi} c_{\rm s}H$ is the effective kinematic viscosity (e.g. \citealt{Matzner&Levin05}). 

\citet{Kratter&Matzner06}, hereafter KM06, examine the conditions under which disks around massive proto-stars fragment during the embedded phase.  In particular they determine at what radii the disk can support accretion at the core-supplied in-fall rate (viz.~$\dot{M}_{\rm c} \lesssim \dot{M}_{\rm max}$), taking into account the stabilizing effects of viscous turbulent heating and irradiation from the central star.  KM06 conclude that for typical core accretion rates $\dot{M}_{c} \sim 10^{-3}-10^{-4}M_{\sun}$ yr$^{-1}$ the disk fragments outside of a critical radius $R_{\rm frag}$, which robustly lies in the range $R_{\rm frag} \sim 100-200$ AU for the entire range in stellar mass $M_{\star} \sim 10-100M_{\sun}$.

KM06 then compare $R_{\rm frag}$ to the typical circularization radius $\bar{R}_{\rm circ}$, which is determined by the angular momentum of the infalling envelope.  They estimate that $\bar{R}_{\rm circ}$ increases from $\sim 100$ AU to $\sim 500$ AU as $M_{\star}$ increases from $\sim 10$ to $\sim 100M_{\sun}$, but with significant scatter at all values of $M_{\star}$ (in contrast to $R_{\rm frag}$, which is approximately constant with $M_{\star}$).  From this they conclude that the outer regions of proto-stellar disks around the majority of massive stars are susceptible to fragmentation, presumably resulting in the formation of one or more proto-stellar companions at $R > R_{\rm frag}$.  

The presence of a massive companion at $\sim 10^{2}-10^{3}$ AU would likely preclude the long-term survival of a proto-stellar disk.  A fraction of massive cores will, however, also form stable disks with $R_{\rm circ} < R_{\rm frag}$ from the low end tail of the core angular momentum distribution,\footnote{In fact, by not sharing their mass with binary companions, such stars may preferentially grow to become the most massive (KM06).} which is broad because the core is supported by stochastic, turbulent processes.  Furthermore, even when fragmentation occurs, it is unlikely to completely suppress accretion.  Fragmenting regions may be susceptible to additional, {\it global} gravitational instabilities (e.g. \citealt{Adams+89}; \citealt{Laughlin+98}), which would cause matter to rapidly accrete to radii $\lesssim R_{\rm frag}$ (i.e. with an effective viscosity $\alpha \sim 1$; e.g.~\citealt{Kratter+08}).  Thus, even if a proto-stellar companion forms outside $R_{\rm frag}$, it may migrate inside $R_{\rm frag}$ during the embedded phase, resulting in a tight binary or even coalescence with the central star (e.g. KM06; \citealt{Krumholz+07}; \citealt{Kratter+08}).  Indeed, massive stars are known to have high binary fractions (e.g.~\citealt{Preibisch+01}; \citealt{Lada06}), with apparent mass-ratio and semi-major axis distributions which suggest that the binary orbits were formed during the embedded phase with small separations $\ll R_{\rm frag} \sim 100$ AU (e.g. \citealt{Pinsonneault&Stanek06}; \citealt{Apai+07}; \citealt{Krumholz&Thompson07}).    

In the rest of this paper we focus on the subsequent post-embedded evolution of the disk, assuming that its mass is initially concentrated near a radius $\sim 100$ AU $\lesssim R_{\rm frag}$.  This choice is motivated by the competition between molecular core angular momentum (which favors large radii) and the requirement not to fragment (which requires radii $< R_{\rm frag}$).  We furthermore neglect the possible effects of stellar companions at large radii.  In doing so, we are implicitly assuming that either (1) the disk forms from relatively low angular momentum material such that no massive stellar companions form; or (2) companions do form, but they subsequently migrate inwards during the embedded phase, resulting in a single star or tight binary with negligible subsequent effect on the disk evolution at larger radii.   

\subsection{Isolated Gravito-Turbulent Phase}
\label{sec:GIphase}

As argued in the previous section, a plausible initial condition following the embedded stage of accretion is a relatively isolated disk with initial radius $R_{d,0} \sim 100$ AU $\lesssim R_{\rm frag}$  and mass $M_{d,0}$ (to be determined below).  Although by construction the disk is stable to fragmentation at its initial radius (i.e. $\Omega t_{\rm cool} \gtrsim 1$), it still resides in a marginally stable state of gravito-turbulence with $Q \sim Q_{0} \sim 1$, which we now characterize.

Using equation (\ref{eq:ToomreQ}), the midplane temperature $T$ of a disk with $Q = Q_{0}$ at the radius $R_{d}$ where the local disk mass $\propto \Sigma R^{2}$ peaks is given by  
\be
T_{\rm gi} \simeq \frac{\mu c_{\rm s}^{2}}{\gamma k} = 450{\rm\,K\,}Q_{0}^{2}A_{2}^{-2}M_{\star,100}(M_{d}/0.1M_{\star})^{2}R_{d,100}^{-1},
\label{eq:T_GI}
\ee 
where $M_{\star} = 100M_{\star,100}M_{\sun}$, $R_{d} \equiv 100R_{d,100}$ AU, $M_{d} \equiv A\pi R_{d}^{2}\Sigma$ is the total disk mass, and $A \equiv 2A_{2}$ is a constant that relates total disk mass to the local properties at $R_{d}$ and which takes the value $A = A_{\rm gi} \approx 2$ for a gravito-turbulent disk in steady state (see Appendix A and the discussion in $\S\ref{sec:ring}$).  The sound speed is $c_{\rm s} = (\gamma kT/\mu)^{1/2}$, where we have taken $\gamma = 7/5$ and $\mu \simeq 2.34m_{\rm H}$ for molecular gas, and $m_{\rm H}$ is the mass of a hydrogen atom.  Likewise, the disk scaleheight at $R \approx R_{d}$ is given by
\be
\left.\frac{H}{R}\right|_{R_{d}} \simeq 0.05Q_{0}(M_{d}/0.1M_{\star})A_{2}^{-1}
\label{eq:hoverr_gi}
\ee   

The massive proto-stellar disks of interest are dusty and optically-thick, with a midplane optical depth $\tau = \kappa\Sigma/2$, where $\kappa$ is the dust opacity.  The radiative flux through the disk surface is $F \approx 8\sigma T^{4}/3\tau$ (assuming a constant heating rate per unit mass), and the cooling time is thus given by \citep{Kratter+10}
\begin{eqnarray}
&t_{\rm cool}& \simeq \frac{3\gamma\Sigma c_{\rm s}^{2}\tau}{32(\gamma-1)\sigma T^{4}}
\nonumber \\
& \approx & 1180{\,\rm yr\,}Q_{0}^{-6}A_{2}^{4}\kappa_{10}M_{\star,100}^{-1}(M_{d}/0.1M_{\star})^{-4}R_{d,100}^{-1},
\label{eq:tcool}
\end{eqnarray}
where we have scaled $\kappa \equiv 10\kappa_{10}$ cm$^{2}$ g$^{-1}$ to a constant value that is representative of the limited opacity range ($\sim 3-16$ cm$^{2}$ g$^{-1}$) appropriate throughout the temperature range $10^{2}{\,\rm K} \lesssim T \lesssim 10^{3}$ K of present interest \citep{Semenov+03}.  The gravito-turbulent viscosity (eq.~[\ref{eq:alphaGI}]) corresponding to the cooling rate in equation (\ref{eq:tcool}) is given by 
\begin{eqnarray}
\alpha_{\rm gi} \approx& 1.2\times 10^{-2}Q_{0}^{6}A_{2}^{-4}\kappa_{10}^{-1}M_{\star,100}^{1/2}(M_{d}/0.1M_{\star})^{4}R_{d,100}^{5/2},
\label{eq:alpha_GI}
\end{eqnarray}
corresponding to a viscous timescale
\begin{eqnarray}
&t_{\rm visc,gi}& \approx \frac{R_{d}^{2}}{\alpha_{\rm gi}c_{\rm s}H} \nonumber \\
& \approx & 0.43 {\,\rm Myr\,}Q_{0}^{-8}A_{2}^{6}\kappa_{10}M_{\star,100}^{-1}(M_{d}/0.1M_{\star})^{-6}R_{d,100}^{-1}\nonumber \\
\label{eq:tvisc_GI}
\end{eqnarray}

A reasonable definition for the end of the embedded phase and the onset of ``isolated'' accretion is when the disk mass has decreased sufficiently that the accretion timescale of the disk $t_{\rm visc} \propto M_{d}^{-6}$ (eq.~[\ref{eq:tvisc_GI}]) exceeds the infall accretion timescale from the progenitor molecular core $t_{\rm c} \sim 10^{5}$ years \citep{McKee&Tan03}.  By equating $t_{\rm visc,gi} = t_{\rm c}$, we thus find that the disk mass at the end of the embedded phase is:
\begin{eqnarray}
\left.\frac{M_{d}}{M_{\star}}\right|_{\rm t_{\rm visc} = t_{\rm c}} \approx 0.13Q_{0}^{-4/3}A_{2}(t_{\rm c}/10^{5}{\,\rm yr})^{-1/6}\kappa_{10}^{1/6}M_{\star,100}^{-1/6}R_{0,100}^{-1/6}. 
\label{eq:mdiskratio_GI}
\end{eqnarray}
Robustly then, the disk mass is $\sim 10-20$ per cent of the stellar mass at the beginning of the disk's isolated evolution.

Once the disk no longer accretes appreciable external mass and angular momentum, its subsequent evolution occurs with an approximately constant total angular momentum $J_{d} \propto M_{d}(GM_{\star}R_{d})^{1/2}$, neglecting the effects of disk outflows.  As matter accretes, $R_{d}$ thus increases $\propto M_{d}^{-2}$, i.e. the disk viscously spreads to larger radii \citep{Pringle81}.  

For times much greater than the initial viscous time $t_{\rm visc,0} \sim 10^{5}$ yrs, the properties of the disk at $R_{d}$ asymptote to a self-similar evolution which is characterized by $t \sim t_{\rm visc}$, where $t_{\rm visc}$ is the viscous time at $R_{d}$ (see $\S\ref{sec:models}$).  For example, from equation (\ref{eq:tvisc_GI}) we infer that $t_{\rm visc} \propto M_{d}^{-6}R_{d}^{-1} \propto M_{d}^{-4}$, implying that $M_{d} \propto t_{\rm visc}^{-1/4} \sim t^{-1/4}$.  As discussed in Appendix A, since gravito-turbulent disks obey a non-linear diffusion equation, it is not clear that their the viscous evolution is indeed controlled by the disk properties at the peak radius $R_{d}$.  Nevertheless, we argue that a self-similar approach may still be justified, which results in the following solution (Appendix B)
\be
M_{d} \simeq 7.9M_{\sun}Q_{0}^{-4/3}\kappa_{10}^{1/6}M_{\star,100}^{5/6}R_{0,100}^{-1/6}(t/t_{\rm visc,0})^{-1/4} 
\label{eq:MdGIss}
\ee
\be
R_{d} \simeq 2.2R_{d,0}(t/t_{\rm visc,0})^{1/2}
\ee
\be
T|_{R_{d}} \simeq 130{\rm K}Q_{0}^{-2/3}\kappa_{10}^{1/3}M_{\star,100}^{2/3}R_{0,100}^{-4/3}(t/t_{\rm visc,0})^{-1} 
\label{eq:TGIss}
\ee
\be
\Sigma|_{R_{d}} \simeq 230{\rm \,g\,cm^{-2}}Q_{0}^{-4/3}\kappa_{10}^{1/6}M_{\star,100}^{5/6}R_{0,100}^{-13/6}(t/t_{\rm visc,0})^{-5/4} 
\ee
\be
\tau|_{R_{d}} \simeq 2300Q_{0}^{-4/3}\kappa_{10}^{7/6}M_{\star,100}^{5/6}R_{0,100}^{-13/6}(t/t_{\rm visc,0})^{-5/4} 
\ee
\be
H/R|_{R_{d}} \simeq 0.04 Q_{0}^{-1/3}\kappa_{10}^{1/6}M_{\star,100}^{-1/6}R_{0,100}^{-1/6}(t/t_{\rm visc,0})^{-1/4}
\ee
\be
(\Omega t_{\rm cool}|_{R_{d}})^{-1} \simeq 0.03Q_{0}^{2/3}\kappa_{10}^{-1/3}M_{\star,100}^{-1/6}R_{0,100}^{11/6}(t/t_{\rm visc,0})^{1/4},
\label{eq:omegatcoolGIss}
\ee
\be
Q|_{R_{d}} = Q_{0}
\ee
where we have used equation (\ref{eq:mdiskratio_GI}) for the initial disk mass.

From the above, note that the disk accretes and spreads slowly in time ($M_{d} \propto t^{-1/4}; R_{d}\propto t^{1/2}$) due to the rapid increase in the viscous time with decreasing disk mass ($t_{\rm visc} \propto M_{d}^{-6}$; eq.~[\ref{eq:tvisc_GI}]).  Also note that $(\Omega t_{\rm cool})^{-1} \propto \alpha_{\rm gi}$ increases with time as $t^{1/4}$, which suggests that the disk may eventually become unstable to fragmentation ($\alpha_{\rm gi} \gtrsim 1$).  However, as shown in the next section, stellar irradiation provides a temperature floor that stabilizes self gravity as the disk mass decreases.  In fact, the self-similar solutions above rarely have time to obtain before the disk becomes irradiation-supported.  

\subsection{Irradiation-Supported Phase}
\label{sec:irrad}

Irradiation from the star during the embedded stage acts to stabilize proto-stellar disks from gravitational instability (e.g.~\citealt{Matzner&Levin05}).  As we now discuss, it also supplies longer term support, effectively shutting off gravito-turbulence soon into the disk's isolated evolution.

High mass stars radiate at a significant fraction $\eta_{\rm edd}$ of the Eddington limit, with a luminosity $L_{\star} = \eta_{\rm edd}L_{\rm edd} = 1.3\times 10^{40}\eta_{\rm edd}M_{\star,100}$ ergs s$^{-1}$.\footnote{Accretion luminosity is generally negligible in comparison to intrinsic stellar luminosity for massive stars.}  Stars of mass $M_{\star} = 25(100)M_{\sun}$, for instance, have $\eta_{\rm edd} = 0.1(0.4)$ upon entering the main sequence (e.g.~\citealt{Maeder&Meynet87}).  Assuming that the disk is flared, such that the scaleheight $H$ increases with radius faster than $\propto R$, the stellar flux incident on the disk normal at radius $R$ is:
\be 
F_{\rm irr} = \frac{Lf}{4\pi R^{2}},
\ee
where $f = \eta_{\rm f}(H/R)$ is the fraction of the total flux intercepted and absorbed by the disk and $\eta_{\rm f} \sim$ few characterizes the precise flaring geometry and number of vertical scaleheights to the disk photosphere \citep{Chiang&Goldreich97}, which depends on details such as grain settling.  In our analytic estimates below we combine $\eta_{\rm edd}$ and $\eta_{\rm f}$ into a single constant $\eta = \eta_{\rm edd}\eta_{\rm f} \sim 1$.  Few of our results depend sensitively on $\eta$.

Irradiation dominates over viscous dissipation in heating the midplane when $F_{\rm irr} \gtrsim F_{\rm visc}/\tau$, where $F_{\rm visc}$ is the heat flux due, in this case, to the nominal level of gravito-turbulence (e.g.~\citealt{Rafikov09}).  When this condition is satisfied the resulting midplane temperature $T_{\rm irr}$ is determined by equating $F_{\rm irr}$ with the vertical radiative flux $F \simeq 8\sigma T^{4}/3$ \citep{Kratter+10}, giving
\be
T_{\rm irr} \simeq 686{\rm\,K\,}\eta^{2/7}M_{\star,100}^{1/7}R_{d,100}^{-3/7},
\label{eq:T_irrad}
\ee
corresponding to a disk scaleheight
\be
\frac{H}{R} = 0.06\eta^{1/7}M_{\star,100}^{-3/7}R_{d,100}^{2/7}
\label{eq:hoverr_irrad}
\ee
The condition that the disk is irradiation-supported and that equations (\ref{eq:T_irrad}) and (\ref{eq:hoverr_irrad}) are valid, namely that $F_{\rm irr} \gtrsim F_{\rm visc}/\tau$, is equivalent to the requirement that $T_{\rm irr}$ exceed $T_{\rm gi}$ (eq.~[\ref{eq:T_GI}]).  This condition is satisfied for disk masses below the critical mass
\be
\left.\frac{M_{d}}{M_{\star}}\right|_{T_{\rm irr} = T_{\rm gi}} = 0.12Q_{0}^{-1}A_{2}\eta^{1/7}M_{\star,100}^{-3/7}R_{d,100}^{2/7}
\label{eq:diskmassratio_irr}
\ee
Note that this mass is comparable to the initial disk mass estimated in equation (\ref{eq:mdiskratio_GI}).  This shows that the isolated gravito-turbulent stage discussed in $\S\ref{sec:GIphase}$ is, at best, short-lived because $T_{\rm gi}$ decreases rapidly $\propto t^{-1}$ during the gravito-turbulent phase (eq.~[\ref{eq:TGIss}]) and irradiation provides a stable temperature floor.  In fact, in some cases $M_{d}$ in equation (\ref{eq:diskmassratio_irr}) exceeds that given in equation (\ref{eq:mdiskratio_GI}), in which case the disk is irradiation-supported from the very onset of its isolated evolution. 

Once the disk midplane becomes supported by external radiation, gravito-turbulent heating is no longer necessary to maintain $Q \gtrsim Q_{\rm 0} \sim 1$.  
Without GI-induced torques, the most viable alternative source of disk viscosity is turbulence generated by the magneto-rotational instability (MRI; \citealt{Balbus&Hawley92}).  If the MRI operates throughout the entire disk, the resulting turbulent shear stress can, for our purposes, be approximately described by a Shakura-Sunyaev $\alpha$-viscosity (e.g. \citealt{Hawley+95}; \citealt{Fromang+07}), with typical values $\alpha \sim 10^{-2}-10^{-1}$ inferred from simulations and observations (e.g.~\citealt{Chiang&Murray-Clay07}; \citealt{King+07}).  The accretion timescale in an irradiation-supported $\alpha$-disk is thus given by  
\begin{eqnarray}
&&t_{\rm visc,irr} = \frac{R_{d}^{2}}{\alpha c_{\rm s}H} \nonumber \\
&& \approx  0.44{\rm\,Myr\,}(\alpha /10^{-2})^{-1}\eta^{-2/7}M_{\star,100}^{5/14}R_{d,100}^{13/14},
\label{eq:tvisc_irr}
\end{eqnarray}
where we have used $H/R$ from equation (\ref{eq:hoverr_irrad}).  Note that for values $\alpha \gtrsim 10^{-2}$ which are expected for fully-ionized disks, the accretion timescale is rather short compared to the lifetimes $t_{\rm life} \sim 3-10$ Myr of very massive stars.  

Proto-stellar disks are, however, dense, cold, and hence weakly ionized.  When the conductivity is sufficiently low, MRI modes are stabilized because the gas collisionally decouples from the magnetic field and ionized species decouple from the neutral molecular hydrogen that comprises the bulk of the disk's mass  (e.g.~\citealt{Blaes&Balbus94}; \citealt{Reyes-Ruiz&Stepinski95}; \citealt{Jin96}).  In particular, the surface column through disk near $R_{d}$ at the beginning of its irradiation-supported evolution is:
\begin{eqnarray}
&&\Sigma_{\rm 0} \simeq \frac{M_{d}|_{T_{\rm irr} = T_{\rm gi}}}{A_{\rm gi}\pi R_{d}^{2}} \approx \nonumber \\
&& 1.7\times 10^{3}{\,\rm g\,cm^{-2}}Q_{0}^{-1}\eta^{1/7}M_{\star,100}^{4/7}R_{d,100}^{-12/7},
\label{eq:sigma0}
\end{eqnarray}
where we have used equation (\ref{eq:diskmassratio_irr}).  This large column shields the midplane from ionizing radiation, perhaps creating a ``dead zone'' near the midplane in which little MRI turbulence is generated (\citealt{Gammie96}; \citealt{Turner+07}).

Determining the presence and extent of the dead zone requires estimating the surface column  $\Sigma_{a}$ that is sufficiently ionized to become MRI active.  Cosmic rays are usually assumed to be the chief ionizing agent at large radii in proto-stellar disks because they penetrate to a significant depth $\Sigma \sim 10^{2}$ g cm$^{-2}$  \citep{Umebayashi&Nakano81}.  However, the regions around very massive stars may be shielded from interstellar cosmic rays by magnetized winds from the star or disk.  This is particularly true because massive stellar winds are more powerful {\it throughout the entire stellar lifetime} than the outflows from low mass stars during their pre-main sequence phases.  

Assuming that a fraction $f_{\rm b} \ll 1$ of the total luminosity of a massive star's wind $L_{\rm w} \simeq (1/2)\dot{M}_{\rm w}v_{\rm w}^{2}$ is carried in the form of Poynting luminosity $L_{P} = (B^{2}c/4\pi)\times (4\pi R^{2})$, then the ratio of the Larmor radius of a cosmic ray proton $R_{\rm L} = E_{\rm cr}/eB$ to the radius $R$ is independent of radius:
\begin{eqnarray}
&&\frac{R_{\rm L}}{R} \simeq 10^{-5}\left(\frac{f_{\rm b}}{10^{-2}}\right)^{-1/2}\times \nonumber \\
&&\left(\frac{E_{\rm cr}}{\rm GeV}\right)\left(\frac{\dot{M}_{\rm w}}{10^{-6}M_{\sun}{\,\rm yr^{-1}}}\right)^{-1/2}\left(\frac{v_{\rm w}}{10^{3}{\,\rm km\,s^{-1}}}\right)^{-1},
\label{eq:rloverr}
\end{eqnarray}
where $E_{\rm cr}$ is the cosmic ray energy, $B$ is the magnetic field, and $v_{\rm w}$ and $\dot{M}_{\rm w} = 4\pi\rho_{\rm w}v_{\rm w}R^{2}$ are the wind velocity and mass-loss rate, respectively.  We have scaled $\dot{M}_{\rm w} \sim 10^{-7}-10^{-5}M_{\sun}$ yr$^{-1}$ and $v_{\rm w} \sim 10^{2.5}-10^{3.5}$ km s$^{-1}$ to typical values for main-sequence and post-main-sequence O star winds (e.g.~\citealt{vanBuren85}).  

Equation (\ref{eq:rloverr}) illustrates that cosmic rays near the peak of the energy spectrum ($E_{\rm cr} \sim$ 10 MeV - GeV) have $R_{\rm L} \ll R$, even for very small values of $f_{\rm b}$.  Interstellar cosmic rays should thus act as a fluid and remain exterior to the termination shock of the wind with the ISM, which is $\gg 10^{3}$ AU for typical values of the $L_{\rm w}$ and the ISM pressure.  Although in principle a modest fraction of cosmic rays (especially those with higher energies) could diffuse upstream to ionize the disk, the fraction that reaches the disk at radii $R \lesssim 10^{2}-10^{3}$ AU is probably small.  Cosmic rays could also be accelerated {\it locally} near the disk, due to e.g. shock acceleration in the stellar wind or disk corona, but the likelihood of this possibility and its relative importance are difficult to assess.

X-rays are another source of ionization in proto-stellar disks (\citealt{Glassgold+97}; \citealt{Igea&Glassgold99}).  O stars are strong sources of soft X-rays \citep{Berghoefer+96}, with typical luminosities $L_{\rm X} \sim 10^{32}-10^{33}$ ergs s$^{-1}$ ($\sim 10^{-6}L_{\star}$; \citealt{Flaccomio+03}), which are a factor $\sim 10^{3}-10^{4}$ larger than the typical X-ray luminosities of T Tauri stars.  However, the X-ray {\it flux} at $R \gtrsim 10^{2}$ AU is similar to that experienced by a T-Tauri disk at a $R \sim $ few AU.  The latter case was analyzed by \citet{Igea&Glassgold99}, who found $\Sigma_{a} \sim 10$ g cm$^{-2}$, relatively independent of details such as the precise X-ray spectrum.  Absent cosmic rays, we thus take $\Sigma_{a} \sim 10$ g cm$^{-2}$ as a fiducial estimate of the active layer in massive proto-stellar disks.  We acknowledge, however, that in general the active depth (and even its precise definition) depends on a variety of complex and interrelated processes, including the precise conditions for the growth of the MRI (\citealt{Fleming+00}; \citealt{Pessah+07}); the abundance of gas phase metals; the size, evolution, and settling of dust grains (e.g.~\citealt{Fromang+02}; \citealt{Bai&Goodman09}); and the presence of turbulent mixing (e.g.~\citealt{Ilgner&Nelson06,Ilgner&Nelson08}; \citealt{Turner+07}).

Because the initial column $\Sigma_{0}$ (eq.~[\ref{eq:sigma0}]) greatly exceeds $\Sigma_{a} \sim 10$ g cm$^{-2}$, proto-stellar disks around massive stars likely possess an extensive dead zone at the onset of the irradiation-supported phase.  Although there have been several studies of [and speculations regarding] the effects of a dead zone on disk angular momentum transport, the present theoretical picture remains largely incomplete \citep{Stone+00}.  Below we compare and contrast three qualitatively distinct possibilities for the viscous evolution of the disk during the irradiation-supported phase.  This discussion motivates the strength and form of the viscosity that we adopt in our disk models in $\S\ref{sec:models}$. 

\subsubsection{Marginal Gravitational Instability}
\label{sec:marginalGI}

One possibility is that angular momentum transport is largely ineffective, such that the accretion timescale greatly exceeds the stellar lifetime $\sim 3-10$ Myr.  In terms of an effective $\alpha-$viscosity (eq.~[\ref{eq:tvisc_irr}]) this requires $\alpha \ll 10^{-3}$.  Under this circumstance, one might expect the disk would not evolve significantly (absent external influences) after it becomes irradiation-supported.  The final mass at stellar core collapse would thus simply be given by equation (\ref{eq:diskmassratio_irr}).  

This analysis neglects, however, the effects of stellar mass loss.  Since the Toomre parameter $Q$ is proportional to the stellar mass (eq.~[\ref{eq:ToomreQ}]), a decreasing value of $M_{\star}(t)$ could drive the disk to again become gravitationally unstable at later stages of stellar evolution.  Because angular momentum transport due to gravitational instabilities is rapid (eq.~[\ref{eq:tvisc_GI}]), the onset of GI would rapidly deplete the disk's mass until it again becomes irradiation-supported.  Stellar mass loss thus locks a disk that otherwise does not evolve into a state of marginal gravitational instability ($\delta Q = 0$), such that the accretion rate is tied to the stellar mass loss rate $\dot{M}_{\star}$ by the relationship
\be
\dot{M}_{d}|_{\delta Q = 0} = (2M_{d}/11M_{\star})\dot{M}_{\star},
\label{eq:mdotmin}
\ee
where we have used equation (\ref{eq:diskmassratio_irr}) and assumed that the disk spreads with constant total angular momentum $J_{d} \propto M_{d}(GM_{\star}R_{d})^{1/2}$.  Thus, in this case the final mass at stellar core collapse is smaller than its initial value at the onset of irradiation-support by a factor $(M_{\star,f}/M_{\star,0})^{2/11}$, where $M_{\star,f} \equiv M_{\star}(t=t_{\rm life})$ is the final stellar mass.  The final radius is $R_{d,f}/R_{d,0} = (M_{\star,f}/M_{\star,0})^{-15/11}$.  Unsurprisingly, negligible non-gravitational viscosity results in the most massive, compact disks which can survive until stellar death.

\subsubsection{Active Zone Draining}
\label{sec:active}

Another possibility is that accretion {\it does} occur in irradiation-supported disks, but it is restricted to regions of the disk which are sufficiently ionized to be MRI active (e.g.~\citealt{Gammie96}; \citealt{Chiang&Murray-Clay07}).  In addition to the externally-ionized surface layer $\Sigma_{a}$ previously discussed, the entire disk is MRI active due to the collisional ionization of alkali metals when the midplane temperature exceeds $T_{a} \approx 10^{3}$ K (\citealt{Gammie96}; \citealt{Fromang+02}).  Using equation (\ref{eq:T_irrad}) this condition is satisfied interior to the radius 
\begin{eqnarray}
R_{a} \equiv R(T=T_{a}) \simeq 40{\,\rm AU\,}(T_{a}/10^{3}{\,\rm K})^{-7/3}\eta^{2/3}M_{\star,100}^{1/3}\nonumber \\
\label{eq:rion}
\end{eqnarray}
From equation (\ref{eq:tvisc_irr}), the accretion timescale at $R_{a}$ is thus \begin{eqnarray}
 t_{\rm visc,irr}|_{R_{a}} \approx 0.02{\rm\,Myr\,}(\alpha/0.1)^{-1}\eta^{1/3}M_{\star,100}^{2/3}(T_{a}/10^{3}{\,\rm K})^{-13/6},\nonumber\\
\label{eq:tvisc_irr_ion}
\end{eqnarray}
where we now scale the viscosity to a larger value $\alpha \sim 0.1$ which is appropriate for the MRI-active inner rim of the disk. 

Although $t_{\rm visc,irr}|_{R_{a}}$ is very short, it is not the timescale for the entire disk to accrete.  At radii $R > R_{a}$ only the active surface column $\Sigma_{a}$ feeds the inner disk and, as a result, the accretion rate reaches a minimum just outside $R_{a}$ \citep{Gammie96}.  The inner edge of the dead zone thus acts like a spigot by controlling the net inflow rate for the entire disk.  This results in an {\it effective} accretion timescale for the disk to ``drain'' through its active layers $t_{\rm drain}$ which is larger than equation (\ref{eq:tvisc_irr_ion}) by the ratio of the total disk mass to the ``active'' mass at $R \approx R_{a}$, $M_{\rm a} \sim \pi R_{a}^{2}\Sigma_{a}$:
\begin{eqnarray}
t_{\rm drain}&& \sim (M_{d}/M_{a})t_{\rm visc,irr}|_{R_{a}} \approx 32{\,\rm Myr\,}(\alpha/0.1)^{-1}\eta^{-1}\times \nonumber \\
&& M_{\star,100}(M_{d}/0.1M_{\star})(T_{a}/10^{3}{\,\rm K})^{5/2}(\Sigma_{\rm a}/10{\,\rm g\,cm^{-2}})^{-1}. \nonumber \\
\label{eq:tvisc_irr_eff}
\end{eqnarray}
For typical values of the disk's mass at the onset of the irradiated phase (eq.~[\ref{eq:diskmassratio_irr}]), $t_{\rm drain}$ generally exceeds the stellar lifetime $t_{\rm life} \sim 3-10$ Myr for $M_{\star,0} \gtrsim 25M_{\sun}$.  This suggests that the disk is unlikely to evolve significantly prior to stellar death if it accretes solely through its active regions; its evolution would thus revert to the case of negligible irradiation-supported viscosity described in $\S\ref{sec:marginalGI}$.  In the next section we discuss the perhaps more plausible situation that limited turbulent angular momentum transport (perhaps seeded in the active layers) acts throughout the {\it entire} disk.

\subsubsection{(Reduced) Global Alpha Viscosity}
\label{sec:alpha}

A final possibility is that proto-stellar disks undergo a viscous evolution that is qualitatively similar to that of fully-ionized disks, but with a lower ``effective'' value of the $\alpha$ parameter (yet larger than the case of negligible viscous evolution described in $\S\ref{sec:marginalGI}$).  This could occur if MRI turbulence is limited to the active surface layer $\Sigma_{a}$, but this still generates a limited Reynolds \citep{Fleming&Stone03} or Maxwell stress \citep{Turner+07} in the midplane.  The simulations of \citet{Fleming&Stone03}, for instance, found that the effective angular momentum transport was reduced by a factor of $\sim 10$ in the case of an active layer column which was $\sim 20\%$ of the total column.  In this case we would predict ``effective'' viscosities $\alpha \sim 10^{-3}-10^{-2}$ assuming ``standard'' (fully-ionized) values $\alpha \sim 10^{-2}-10^{-1}$, at least when the disk surface density obeys $\Sigma|_{R_{d}} \gg \Sigma_{a}$.  In general we might expect $\alpha$ to be a decreasing function of $\Sigma/\Sigma_{\rm a}$ (which asymptotes to the fully ionized value of $\alpha$ as $\Sigma \rightarrow 2\Sigma_{\rm a}$), but since this dependence has yet to be determined we hereafter assume $\alpha = $ constant.  

Assuming a constant $\alpha$ disk model with an initial mass as given in equation (\ref{eq:diskmassratio_irr}), the late-time similarity solutions (analogous to eqs.~[\ref{eq:MdGIss}]$-$[\ref{eq:omegatcoolGIss}]; see Appendix B) are given by
\be
M_{d} \simeq 14M_{\sun}Q_{0}^{-1}\eta^{1/7}M_{\star,100}^{4/7}R_{0,100}^{-2/7}(t/t_{\rm visc,0})^{-7/13}
\label{eq:Mdirrss}
 \ee
\be
R_{d} \simeq 1.7R_{d,0}(t/t_{\rm visc,0})^{14/13} \ee
\be
T|_{R_{d}} \simeq 550{\rm K}\eta^{2/7}M_{\star,100}^{1/7}R_{0,100}^{-3/7}(t/t_{\rm visc,0})^{-6/13} \ee
\be
\Sigma|_{R_{d}} \simeq 200{\,\rm g\,cm^{-2}}Q_{0}^{-1}\kappa_{10}M_{\star,100}^{4/7}R_{0,100}^{-12/7}(t/t_{\rm visc,0})^{-35/13} \ee
\be
\tau|_{R_{d}} \simeq 2000Q_{0}^{-1}\kappa_{10}M_{\star,100}^{4/7}R_{0,100}^{-12/7}(t/t_{\rm visc,0})^{-35/13} \ee
\be
H/R|_{R_{d}} \simeq 0.07 \eta^{1/7}M_{\star,100}^{-3/7}R_{0,100}^{2/7}(t/t_{\rm visc,0})^{4/13}\ee
\be
Q|_{R_{d}} \simeq 0.5Q_{0}(t/t_{\rm visc,0})^{11/13},
\label{eq:Qirrss}
\ee
where $t_{\rm visc,0}$ here refers to equation (\ref{eq:tvisc_irr}) evaluated at the beginning of the irradiation-supported phase, and we have taken $A = A_{\rm irr} \simeq 6.8$, as appropriate for an irradiation-supported $\alpha-$disk in steady-state (see Appendix A).  These solutions are analogous to those obtained by \citet{Hartmann+98} as applied to T Tauri disks.

From the above note the following: (1) the relatively slow evolution of the disk $M_{d} \propto t^{-7/13}$ implies that the disk can have substantial mass on timescales $\gg t_{\rm visc,0}$; (2) our assumptions that the midplane is optically thick and that the temperature remains sufficiently high for our assumed opacity ($\gtrsim 100$ K) remains valid for several viscous times; (3) $Q$ increases with time, implying that the disk evolves to an increasingly gravitationally-stable state (i.e. away from the marginally unstable state described in $\S\ref{sec:marginalGI}$); (4) the disk column $\Sigma$ remains $\gtrsim \Sigma_{\rm a} \sim 10$ g cm$^{-2}$ for at least a few viscous times, thereby justifying our use of a ``reduced'' value for the viscosity $\alpha$ on timescales $\gtrsim t_{\rm visc,0}$.

\section{Disk Evolution Models}
\label{sec:models}

We now combine the results and intuition gained from the previous section to construct a simple model for the isolated evolution of massive proto-stellar disks following the embedded stage.  Although several detailed time-dependent models of proto-stellar disks have been constructed by previous authors (e.g. \citealt{Clarke+01}; \citealt{Matsuyama+03}; \citealt{Alexander+06}; \citealt{Zhu+09}; \citealt{Gorti&Hollenbach09b}), our calculations are unique in focusing on the evolution of disks around very massive stars for the entirety of the stellar lifetime.  The time-dependent disk properties we obtain are used to assess the effects of external, destructive influences in $\S\ref{sec:destruct}$.   

\subsection{Self-Similar Ring Model}
\label{sec:ring}

Our disk model closely follows that used by \citet{Metzger+08} to study accretion following compact object mergers (see e.g. \citealt{Cannizzo+90} and \citealt{Hartmann+98} for related self-similar models).  

At a given time $t$, the disk can be divided into three regions, depending on the local viscous time $t_{\rm visc}$, which generally increases with radius for a fixed total angular momentum (e.g. eq.~[\ref{eq:tvisc_irr}]).  At small radii, $t_{\rm visc} < t$, and the disk enters a steady state with $\dot{M} \propto \nu\Sigma$ constant, where $\nu$ is the kinematic viscosity.  Larger radii where $t_{\rm visc} \sim t$ contain the majority of the disk's mass and angular momentum.  This region determines the viscous evolution of the rest of the disk, including the mass accretion rate that is fed into the interior, steady-state region.  Exterior to this point is a region with $t_{\rm visc} > t$, but this contains only a small fraction of the mass and does not significantly affect the viscous evolution (see, however, the discussion of gravito-turbulent disks in Appendix A).  

Our model treats the disk as a single annulus (the ``ring'') that is evolved forward in time.  The properties of the ring, such as the surface density $\Sigma$ and temperature $T$, are representative of the location $R \simeq R_{d}$ where the local mass $\Sigma r^{2}$ peaks.  The time evolution of the disk is determined by the conservation equations for mass and angular momentum:
\be
	\frac{d}{dt}\left(A\pi\Sigma R_{d}^2\right)=-\dot{M}_d,
	\label{eq:mass}
\ee  
\be 
\frac{d}{dt}\left[B(GM_{\star}R_{d})^{1/2}\pi\Sigma R_{d}^2\right]=-\dot{J} = 0,
	\label{eq:ang_mom}
\ee
where $\dot{M}_d$ is the mass accretion rate, $\dot{J}$ is the angular momentum loss rate, which we take to be zero (i.e. we neglect disk outflows), and $A$ and $B$ are constants which relate the total disk mass and angular momentum to their local values near $R_{d}$.   At early times the constants $A$ and $B$ depend on how matter is initially spatially distributed.  At times much greater than the initial viscous time, material spread in a manner determined by the viscosity.  As described in Appendix A, $A$ and $B$ are determined by setting the solution of our simplified ring model at late times equal to the solution for a spreading ring.  In the irradiation-supported accretion phase  ($\S\ref{sec:irrad}$) the viscosity obeys $\nu \propto r^{15/14}$ and this analysis results in exact values ($A_{\rm irr},B_{\rm irr}$) = (6.80,5.85).  In the gravito-turbulence phase ($\S\ref{sec:GIphase}$) the situation is more complicated; since the viscosity obeys $\nu \propto \Sigma^{6}r^{15}$, the surface density evolves via a non-linear diffusion equation which concentrates the mass approximately equally per unit decade in radius (see Fig.~\ref{fig:diskevo}).  As described in Appendix A, we take $A_{\rm gi} \approx 2$ in this case. 

The accretion rate depends on the characteristic mass and viscous timescale of the ring as
\be
	\dot{M}_{d} = fM_d/t_{\rm visc},
	\label{eq:mdotacc}
\ee where the factor $f$ is set, like $A$ and $B$, to match the solution of a spreading ring (Appendix A) and has the values $f_{\rm gi} \simeq 3$ and $f_{\rm irr} \simeq 1.4$ for the gravito-turbulent and irradiation-supported phases, respectively.  The viscous time $t_{\rm visc}$ is determined using equation (\ref{eq:tvisc_GI}) when the disk is gravito-turbulent (i.e. when $T_{\rm gi} > T_{\rm irr}$; eq.~[\ref{eq:tvisc_irr}]).  When the disk is irradiation-supported (i.e. $T_{\rm irr} > T_{\rm gi}$) we use equation (\ref{eq:tvisc_irr}) assuming a  fixed value $\alpha \sim 10^{-3}-10^{-2}$, as motivated by the discussions in $\S\ref{sec:alpha}$ and $\S\ref{sec:active}$.

Equations ($\ref{eq:mass}$) and ($\ref{eq:ang_mom}$) provide two coupled equations to be solved for the dependent variables $R_{d}$ and $M_{d} = A\pi R_{d}^{2}\Sigma$ given $R_{d}(0)$, $M_{d}(0)$ and $M_{\star}(t)$.  The initial conditions reflect those at the end of the embedded phase.  In particular, we take the initial disk mass as the minimum of equations (\ref{eq:mdiskratio_GI}) and (\ref{eq:diskmassratio_irr}), assuming a value for the critical Toomre stability parameter $Q_{0} = 1$.  In all of our models we take an initial radius $R_{d,0} = 100$ AU, as motivated by the work of KM06 and the discussion in $\S\ref{sec:embed}$.  We smoothly interpolate the parameters $y = A$, $B$, and $f$ between the gravito-turbulent (gi) and irradiation-supported (irr) accretion regimes by employing the simple prescription 
\be 
y = y_{\rm irr}\exp[-T_{\rm gi}/T_{\rm irr}] + y_{\rm gi}(1-\exp[-T_{\rm gi}/T_{\rm irr}]) 
\ee
For all stellar masses we take $t_{\rm c} = 10^{5}$ years for the duration of the embedded phase, although our results are insensitive to this choice.  We run the calculation until the end of the stellar lifetime ($t = t_{\rm life}$).

Stellar evolution models show that wind mass-loss is important throughout the main sequence and post main sequence evolution of very massive stars (e.g.~\citealt{Maeder&Meynet87}).  Our models take into account the effects of a decreasing stellar mass $M_{\star}(t)$, both through its affect on disk properties (e.g. $t_{\rm visc}$) and in the angular momentum equation  (eq.~[\ref{eq:ang_mom}]).  In most of our models we use $M_{\star}(t)$ and $L_{\star}(t)$ from \citet{Maeder&Meynet87}, but we also perform comparision calculations in which we neglect mass loss entirely, as may be justified for stars with low metallicity or if the mass loss rates employed in standard stellar evolutionary models are overestimated due to e.g. wind clumping.  However, for simplicity we neglect the resulting shortening of the stellar lifetime.

\subsection{Results}

We calculate evolutionary disk models for stars with initial (post-embedded) masses $M_{\star} = 25-120M_{\sun}$ and for different values of the viscosity $\alpha$ during the irradiation-supported phase.  Table \ref{table:death} summarizes the properties of the disk at stellar core collapse ($t = t_{\rm life}$).  For the case of very low $\alpha$ ($\ll 10^{-3}$), the disk evolves in a state of marginal gravitational instability, as described in $\S\ref{sec:marginalGI}$.  As our results in Table \ref{table:death} confirm, this results in the most massive, compact disks at core collapse.  

Sample solutions for the case of finite values of $\alpha$ are shown in Figures \ref{fig:25msun} and \ref{fig:85msun}.  As we now describe, these cases highlight important differences between the evolution of disks around less massive and more massive stars.

\subsubsection{$M_{\star,0} = 25M_{\sun}$}

\begin{figure}
\resizebox{\hsize}{!}{\includegraphics[ ]{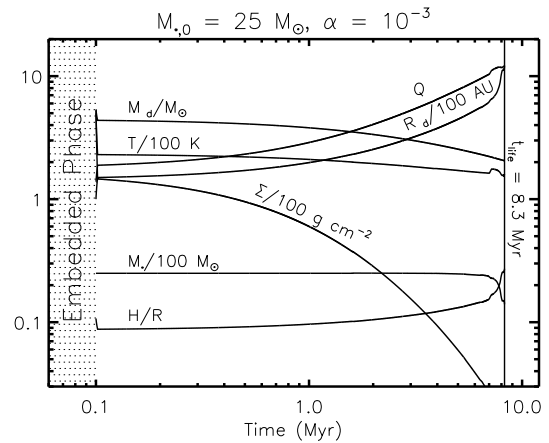}}
\caption{Time evolution of the disk properties for a star with initial mass $M_{\star,0} = 25M_{\sun}$, calculated assuming a value $\alpha = 10^{-3}$ for the viscosity during the irradiation-supported phase.  The disk properties shown include the total mass $M_{d}$ and the peak radius $R_{d}$; and the surface density $\Sigma$, midplane temperature $T$, Toomre parameter $Q$, and the scaleheight $H/R$, all evaluated at the radius $R_{d}$.  Also shown is the stellar mass $M_{\star}(t)$ from the stellar evolution calculations of \citet{Maeder&Meynet87}. }
\label{fig:25msun}

\end{figure}

Figure \ref{fig:25msun} shows the disk properties at the peak radius $R_{d}$ as a function of time for a star with initial mass $M_{\star,0} = 25M_{\sun}$ for an assumed viscosity $\alpha = 10^{-3}$ during the irradiation-supported phase.  As expected, the disk experiences a very brief gravitationally unstable phase of rapid evolution ($\S\ref{sec:GIphase}$), after which it becomes irradiation-supported and its viscous evolution slows considerably ($\S\ref{sec:irrad}$).  The disk mass at the beginning of the isolated phase is $\sim 5M_{\sun}$ and this decreases to $\sim 2M_{\sun}$ by core collapse at $t = t_{\rm life} = 8.3$ Myr.  As a result of angular momentum conservation, the disk spreads to $\sim 1200$ AU by $t = t_{\rm life}$, with a significant fraction of this expansion occurring in just the last $\sim 10^{6}$ years, when the stellar mass decreases to $\sim 14M_{\sun}$ due to stellar winds.  This late spreading occurs even absent significant accretion because conservation of total angular momentum requires that $R_{d}$ increase $\propto M_{\star}^{-1}$ as $M_{\star}$ decreases, even at fixed disk mass.  Indeed, in the otherwise identical model neglecting stellar mass loss, the final radius is only $\approx 700$ AU (see Table \ref{table:death}). 

We also note that the surface density $\Sigma$ decreases below $10$ g cm$^{-2}$ for $t \gtrsim 4$ Myr.  Since this is similar to the surface layer $\Sigma_{a}$ which is MRI active due to X-ray ionization ($\S\ref{sec:irrad}$), the dead zone in the disk midplane may be eliminated.  A larger value of $\alpha$ than the fiducial value $10^{-3}$ may thus be more physical at these late times.  A model calculated with $\alpha = 10^{-2}$, however, results in a much larger final radius $\sim 10^{4}$ AU (Table \ref{table:death}).  As we discuss in $\S\ref{sec:destruct}$, such an extended disk is particularly susceptible to photo-ionization mass-loss and stellar collisions.  This illustrates that massive disks may be unlikely to survive around 25$M_{\sun}$ stars, a result in agreement with other work (e.g.~\citealt{Gorti&Hollenbach09b}).  

\subsubsection{$M_{\star,0} = 85M_{\sun}$}

\begin{figure}
\resizebox{\hsize}{!}{\includegraphics[ ]{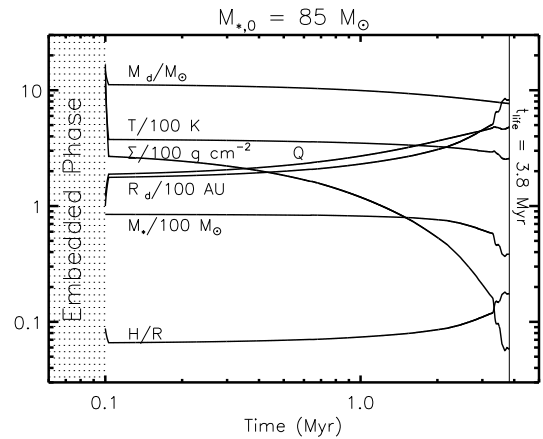}}
\caption{Same as Figure \ref{fig:25msun}, except calculated for an initial stellar mass $M_{\star,0} = 85M_{\sun}$.}
\label{fig:85msun}
\end{figure}

Figure \ref{fig:85msun} shows the disk properties for an initial stellar mass $M_{\star} = 85M_{\sun}$ and $\alpha = 10^{-3}$.  In this case the disk mass at the beginning of the isolated phase is $\approx 17 M_{\sun}$, which decreases to $\approx 8 M_{\sun}$ by core collapse.  Importantly, the surface density remains $\gtrsim 10$ g cm$^{-2} \sim \Sigma_{a}$ throughout almost the entire stellar lifetime (as opposed to the $M_{\star} = 25M_{\sun}$ case), despite significant spreading due to stellar mass loss.  A dead zone may thus be present until stellar core collapse, implying that the ``reduced'' value of the viscosity that we have adopted is self-consistent.  

Given that more massive stars have shorter lifetimes and longer-lived dead zones, we conclude that they likely possess more massive, compact disks at core collapse when considering isolated accretion alone.  



\begin{table}
\begin{center}
\vspace{0.05 in}\caption{Properties of Relic Proto-Stellar Disks at Core Collapse}
\label{table:death}
\resizebox{8cm}{!}{
\begin{tabular}{lccccccccc}
\hline
\hline
\multicolumn{1}{c}{$M_{\star,0}^{(a)}$} &
\multicolumn{1}{c}{$t_{\rm life}^{(b)}$} &
\multicolumn{1}{c}{$\alpha^{(c)}$} &
\multicolumn{1}{c}{$M_{\star,f}^{(d)}$} &
\multicolumn{1}{c}{$M_{d}$} &
\multicolumn{1}{c}{$R_{d}$} &
\multicolumn{1}{c}{$H/R|_{R_{d}}$} &
\multicolumn{1}{c}{$\Sigma|_{R_{d}}$} &
\\
($M_{\sun}$) & (Myr) & & ($M_{\sun}$) & ($M_{\sun}$) & (AU) & & (g cm$^{-2}$) \\
\hline 
\\
120$^{\dagger}$ & 3.5 & $\sim 0$& 64.0 & 13.3 & 240 & 0.09 & 330 \\
120 & - & $10^{-3}$ & - & 10.4 & 670 & 0.14 & 12 \\
120$^{*}$ & - & $10^{-3}$ & - & 10.2 & 380 & 0.09 & 35 \\
120 & - & $10^{-2}$ & - & 4.4 & 3800 & 0.22 & 0.13 \\
120$^{*}$ & - & $10^{-2}$ & - & 4.2 & 2300 & 0.15 & 0.34 \\
85$^{\dagger}$ & 3.8 & $\sim 0$ & 38.4 & 9.4 & 300 & 0.12 & 150 \\
85 & - & $10^{-3}$ & - & 7.7 & 830 & 0.18 & 5.8 \\
85$^{*}$ & - & $10^{-3}$ & - & 7.4 & 400 & 0.10 & 22 \\
85 & - & $10^{-2}$ & - & 3.1 & 5200 & 0.30 & 0.05 \\
85$^{*}$ & - & $10^{-2}$ & - & 2.9 & 2700 & 0.17 & 0.17 \\
60$^{\dagger}$ & 5.4 & $\sim 0$ & 21.3 & 7.4 & 410 & 0.18 & 62 \\
60 & - & $10^{-3}$ & - & 5.4 & 1140 & 0.24 & 2.2 \\ 
60$^{*}$ & 5.4 & $10^{-3}$ & - & 5.3 & 440 & 0.12 & 13\\ 
60 & - & $10^{-2}$ & - & 2.1 & 8000 & 0.42 & 0.014 \\
40$^{\dagger}$ & 5.4 & $\sim 0$ & 9.8 & 5.5 & 680 & 0.29 & 17 \\
40 & - & $10^{-3}$ & - & 3.6 & 1900 & 0.37 & 0.5 \\ 
40$^{*}$ & 5.4 & $10^{-3}$ & - & 3.5 & 500 & 0.14 & 6.0\\ 
40 & - & $10^{-2}$ & - & 1.3 & 13000 & 0.67 & 0.002 \\
25$^{\dagger}$ & 8.3 & $\sim 0$ & 14.0 & 4.9 & 220 & 0.17 & 140 \\
25 & - & $10^{-3}$ & - & 2.1 & 1200 & 0.27 & 0.64 \\
25$^{*}$ & - & $10^{-3}$ & - & 2.0 & 700 & 0.18 & 1.8 \\
25 & - & $10^{-2}$ & - & 0.70 & 11000 & 0.51 & 0.002 \\

\hline
\end{tabular}
}
\end{center}
{\small
$^{(a)}$ Initial (post-embedded) stellar mass; $^{(b)}$ Stellar lifetime; $^{(c)}$ Viscosity during the irradiation-supported phase; $^{(d)}$ Stellar mass at core collapse (from the evolutionary models of \citealt{Maeder&Meynet87}); $^{\dagger}$Calculated assuming negligible angular momentum transport during the irradiation-supported phase (see $\S\ref{sec:marginalGI}$) $^{*}$Calculated assuming negligible stellar mass loss.}
\end{table}

\section{Disk Dispersal Processes}
\label{sec:destruct}

Our results in $\S\ref{sec:models}$ show that isolated evolution generally results in a relatively massive proto-stellar disk at core collapse.  However, it is well known that accretion alone cannot explain the observed lifetimes of disks around lower mass proto-stars.  In this section we thus address {\it external} processes that may act to disperse the disk.  These include photo-evaporation (\S\ref{sec:photo}), stellar collisions (\S\ref{sec:collision}), stripping by continuous stellar winds (\S\ref{sec:wind}), and explosive (LBV-like) stellar eruptions (\S\ref{sec:LBV}).  In $\S\ref{sec:summary}$ we combine our conclusions to assess whether and under what conditions a proto-stellar disk is most likely to survive until core collapse.

\subsection{Photo-evaporation}
\label{sec:photo}

Perhaps the most important means of disk dispersal is photo-evaporation due to UV irradiation (\citealt{Hollenbach+94}; \citealt{Shu+93}; \citealt{Clarke+01}; \citealt{Gorti&Hollenbach09}).  Very massive stars have large Lyman continuum photon luminosities of $\Phi_{\rm i} \sim 10^{48-49}$ s$^{-1}$, which ionize and heat the upper surface layers of the disk to a temperature $\sim 10^{4}$ K, somewhat analogous to an H II region.  This heating drives a Parker-like thermal wind from the disk's surface.  
A critical radius for photo-evaporation $R_{g}$ occurs where the escape speed of the disk equals the sound speed of $\sim 10^{4}$ K gas, $c_{\rm s}(T = 10^{4}$K) $\equiv c_{\rm s,g} \approx 10$ km s$^{-1}$:
\be
R_{\rm g} = (2GM_{\star}/c_{\rm s,g}^{2}) \simeq 1800{\,\rm AU\,}M_{\star,100}
\label{eq:rg}
\ee
\citet{Hollenbach+94} show that the total photo-evaporation rate is dominated by outflows from radii near $R_{\rm g}$.  For disks that extend to $R \gtrsim R_{\rm g}$ they find that the total mass loss rate is\footnote{Since the outflows from massive stars obey the ``strong wind'' condition of \citet{Hollenbach+94}, the mass loss rate in equation (\ref{eq:mdotph}) formally applies only to disks with outer radii $R_{d} \sim R_{\rm g}$.  In the present situation this distinction is, however, not essential; because the evaporation timescale $t_{\rm ph}$ (eq.~[\ref{eq:tph}]) is generally shorter than viscous spreading time, photo-evaporation will truncate the outer edge of any disk that expands to radii $\gtrsim R_{\rm g}$.}
\begin{eqnarray}
\dot{M}_{\rm ph} \simeq 4\times 10^{-5}M_{\sun}{\,\rm yr^{-1}}({\Phi_{\rm i}/10^{49}{\,\rm s^{-1}}})^{1/2}M_{\star,100}^{1/2},
\label{eq:mdotph}
\end{eqnarray}
resulting in an evaporation timescale $t_{\rm ph}$ given by 
\begin{eqnarray}
&&t_{\rm ph} \equiv M_{d}/\dot{M}_{\rm ph} \approx \nonumber \\ 
&&3\times 10^{5}{\rm \,yr}(M_{d}/0.1M_{\star})M_{\star,100}^{1/2}(\Phi_{\rm i}/10^{49}{\,\rm s^{-1}})^{-1/2}.
\label{eq:tph}
\end{eqnarray}
Since $t_{\rm ph}$ is generally $\ll$ the stellar lifetime $t_{\rm life} \sim 3-10$ Myr, this implies that if a disk spreads to radii $\gtrsim R_{\rm g}$, it will photo-evaporate prior to core collapse.  

At radii less than $R_{\rm g}$, however, the mass-loss rate is highly suppressed ($\dot{M}_{\rm ph} \propto \exp[-R/2R_{\rm g}]$, approximately) because the sonic radius $\sim R_{g}$ occurs far out of the midplane, many scale-heights down the disk's exponential atmosphere (e.g. \citealt{Adams+04}).  Disks with outer radii $R_{d}$ less than a few times $R_{g}$ thus have $t_{\rm ph} > t_{\rm life}$ and should not photo-evaporate.

For $M_{\star} = 25M_{\sun}$ the photo-evaporation radius $R_{g}$ is only $\approx 500$ AU.  This is less than or comparable to the size of the disk at core collapse that we found for $M_{\star} = 25M_{\sun}$ in $\S\ref{sec:models}$ (Fig.~\ref{fig:25msun}; Table \ref{table:death}), even in the models that neglect stellar mass loss.  Photo-evaporation may thus be effective at dispersing the proto-stellar disks of $25M_{\sun}$ stars.  

By contrast, for $M_{\star} = 85M_{\sun}$ the critical radius $R_{\rm g} \approx 1500$ AU is significantly larger than the disk radius at core collapse for $\alpha \lesssim 10^{-3}$ (Fig.~\ref{fig:85msun}).  While it is true that $R_{\rm g} \propto M_{\star}$ may decrease significantly due to stellar mass loss, most of this occurs in just the last $\sim 10^{5}$ years before stellar death.  Although the photo-evaporative mass-loss rate from the disk may thus become large (e.g. $\sim 10^{-5}M_{\sun}$ yr$^{-1}$) prior to core collapse, there is insufficient time to photo-evaporate the disk.

To summarize, although photo-evaporation may disperse protostellar disks around less massive stars, higher mass stars become increasingly difficult to photo-evaporate because their deep gravitational potentials imply large values of $R_{\rm g}$ relative to the final disk radius $R_{d}$.  This statement holds true best for disks with small initial radii (as form from low angular momentum cores) and for lower stellar mass-loss rates, as would preferentially occur for low metallicity.  Nevertheless, we emphasize that significant photo-evaporative mass-loss from the disk likely does occur and this may have observable consequences for the supernova (see $\S\ref{sec:type2n}$) and the appearance of its progenitor (\S\ref{sec:disks}). 

\subsection{Stellar Encounters}
\label{sec:collision}

Massive stars generally form in dense stellar clusters (e.g. \citealt{deWit+05}; \citealt{Tan07}).  Close stellar encounters are thus common and may periodically strip off the outer edge of the proto-stellar disk (e.g. \citealt{Clarke&Pringle93}; \citealt{Heller95}; \citealt{Hall+96}).  Although the most massive clusters in the Milky Way have central stellar densities $n_{\star} \sim 10^{4}-10^{5}$ pc$^{-3}$ (e.g.~\citealt{McCaughrean&Stauffer94}; \citealt{Garmire+00}), the density around a {\it typical} O star throughout its lifetime depends on uncertain details such as the IMF (\citealt{Adams&Myers01}; \citealt{Massi+06}) and the degree of primordial (e.g. \citealt{Huff&Stahler06}) and dynamical \citep{PortegiesZwart+04} cluster mass segregation.

In the limit of strong gravitational focusing, the collision timescale for stars with density $n_{\star}$ and one-dimensional velocity dispersion $\sigma$ with a disk of cross section $\pi R_{d}^{2}$ is given by \citep{Binney&Tremaine87}:  
\be
t_{\rm coll} = 4\times 10^{3}{\,\rm yr\,}(n_{\star}/10^{4}{\rm pc^{-3}})^{-1}R_{\rm d,100}^{-1}(\sigma/{\rm 3 km\,s^{-1}})^{-1}M_{\star,100}^{-1}.
\label{eq:tcoll}
\ee
Although this timescale is very short, {\it most} collisions occur with stars near the peak of the IMF, which have a typical mass $M_{\star,0} \sim 0.5M_{\sun}$.  Yet, because incoming stars are typically on near-parabolic [zero-energy] orbits (e.g.~\citealt{Ostriker94}), by energy conservation alone only stars with masses $\gg M_{d}$ are capable of unbinding a significant fraction of the disk.\footnote{For example, in the extreme case that a star of mass $M_{\star}$ is captured into a circular orbit at radius $R_{d}$, only a disk mass $M_{d} = M_{\star}$ could become unbound due to the resulting energy release.}  For most impact trajectories the disk mass removed will be significantly less than that of the colliding star (e.g.~\citealt{Heller95}).  

We may crudely estimate the timescale $t_{\rm se}$ for the disk to be depleted by stellar collisions if we assume that collisions with stars of mass $M_{\star,c}$ dominate the mass loss and that the average mass ejected in a collision is a fraction $\eta_{\rm se}$ of $M_{\star,c}$.  We accomplish this by replacing the stellar density $n_{\star}$ in equation (\ref{eq:tcoll}) with just the fraction $(M_{\star,\rm min}/M_{\star,c})^{1.3}$ of the total density $n_{\star,\rm tot}$ in stars with mass $M_{\star} \sim M_{\star,c}$, where $M_{\star,\rm min} \approx 0.5M_{\sun}$ and we assume a Salpeter IMF $dN_{\star}/dM_{\star} \propto M_{\star}^{-2.3}$ above $M_{\star,\rm min}$.  This gives a depletion timescale 
\begin{eqnarray}
&&t_{\rm se} = t_{\rm coll}|_{n_{\star}=n_{\star,\rm tot}}\times \left(\frac{M_{\star,\rm c}}{M_{\star,\rm min}}\right)^{1.3}\times \frac{M_{d}}{\eta M_{\star,c}} \nonumber \\
&&\sim 1{\,\rm Myr\,}\left(\frac{\eta_{\rm se}}{0.1}\right)^{-1}\left(\frac{M_{d}}{0.1M_{\star}}\right)\left(\frac{M_{\star,\rm c}}{M_{\star,\rm min}}\right)^{0.3}\times \nonumber \\
&&\left(\frac{n_{\star,\rm tot}}{10^{4}{\rm\,pc^{-3}}}\right)^{-1}\left(\frac{R_{d}}{100{\rm\,AU}}\right)^{-1}\left(\frac{\sigma}{3{\rm\,km\,s^{-1}}}\right)^{-1}.
\label{eq:tse}
\end{eqnarray}
We have normalized $\eta_{se}$ to a value $0.1$ which has been conservatively extrapolated from numerical simulations (e.g.~\citealt{Heller95}) to the case in which the disk mass $M_{d}$ and colliding stellar mass $M_{\star,\rm c}$ are comparable.  Simulations suggest that $\eta$ decreases moving to larger ratios $M_{d}/M_{\star,\rm c}$, but to our knowledge no calculations of the disk mass loss have yet been performed in the regime $M_{\star,\rm c} \gtrsim M_{d}$.  

Equation (\ref{eq:tse}) shows that $t_{\rm se}$ may be comparable to the stellar lifetime for disks with masses $M_{d} \sim 0.1M_{\star}$ and radii $R_{d} \sim 100$ AU in clusters with $n_{\star} \lesssim 10^{4}$ pc$^{-3}$.  Though this estimate is extremely crude and the importance of collisions depends sensitively on the conditions specific to an individual massive star, it appears plausible that massive disks could survive until core collapse in ``typical'' cluster environments.  We note, however, the obvious fact that than since $t_{\rm se} \propto R_{d}^{-1}$ more extended disks become increasingly susceptible to dispersal.

\subsection{Stellar Wind Stripping}
\label{sec:wind}

Mass loss due to interaction with the stellar wind is another way to disperse the disk (\citealt{Elmegreen78}; \citealt{Yorke04}; \citealt{Matsuyama+09}).  In evaluating this possibility we closely follow the results of \citet{Matsuyama+09}, hereafter M09, who consider a model in which the central wind obliquely strikes the flared disk surface, driving material out of the system by entraining it in a mixing layer along the disk surface.  M09 estimate that the timescale for significant mass-loss from the disk due to stellar wind stripping is given by (cf.~\citealt{Hollenbach+00})
\begin{eqnarray}
&& \tau_{\rm ws} \equiv \left.\frac{\Sigma}{\dot{\Sigma}}\right|_{R = R_{d}} \sim \frac{M_{d}\cos\beta c_{\rm s}(1+\xi^{2})}{\dot{M}_{\rm w}v_{\rm w}\xi\sin^{2}\gamma} \nonumber \\
&&\approx 10{\,\rm Myr\,}\left(\frac{\xi}{0.1}\right)^{-1}\left(\frac{c_{\rm s}}{{\rm 1 \,km\,s^{-1}}}\right)\left(\frac{M_{d}}{0.1M_{\star}}\right)\left(\frac{M_{\star}}{100M_{\sun}}\right)\times \nonumber \\
&&\left(\frac{v_{\rm w}}{10^{3}{\rm km\,s^{-1}}}\right)^{-1}\left(\frac{\dot{M}_{\rm w}}{10^{-5}M_{\sun}{\rm\, yr^{-1}}}\right)^{-1}\left(\frac{\sin^{2}\gamma}{10^{-2}}\right)^{-1},\nonumber \\
\label{eq:tauws}
\end{eqnarray}
where $\xi \lesssim 0.1$ is the Mach number of the disk material that mixes into the shear layer \citep{Canto&Raga91}, $c_{\rm s}$ is the sound speed in the disk below the mixing layer, and $\beta$/$\gamma$ are the angles that the mixing layer makes with the midplane and the wind, respectively  (see Fig.~1 of M09 for the relevant geometry).  In deriving Equation (\ref{eq:tauws}), the disk density $\rho$ used in determining the mass-flux into the mixing layer $\dot{\Sigma} \propto \rho\xi c_{s}$ is obtained by equating the incident wind ram pressure $\propto \dot{M}_{\rm w}v_{\rm w}\sin^{2}\gamma R_{d}^{-2}$ with the disk thermal pressure $\propto \rho c_{s}^{2}$. 

In the second and third line in equation (\ref{eq:tauws}) we have assumed that $\xi \ll 1$ and $\cos\beta \sim 1$, and we scale $\sin^{2}\gamma$ to the value $\sim 10^{-2}$, motivated by the modest wind-interface incidence angles $\gamma \sim 1-5^{\circ}$ found by M09 as a part of self-consistent numerical calculations (see their Fig.~5).  Although M09 consider lower mass loss rates and velocities than those appropriate for massive stellar winds, they find that $\gamma$ is relatively insensitive to the wind properties (if anything, $\gamma$ appears to decrease with the wind power; see their Fig.~6).

The midplane temperature in the disk ($T \gtrsim 100$ K) sets a lower limit to the sound speed $c_{s} \sim 1$ km s$^{-1}$.  Thus, for properties typical of O star winds, $\dot{M}_{\rm w} \sim 10^{-7}-10^{-5}M_{\sun}$ yr$^{-1}$ and $v_{\rm w} \sim 10^{2.5}-10^{3.5}$ km s$^{-1}$, equation (\ref{eq:tauws}) shows that $\tau_{\rm ws}$ generally exceeds the stellar lifetime $t_{\rm life} \sim 3-10$ Myr.  The M09 model does not consider the possible entrainment of matter from the inner rim of the disk at $R \sim R_{a}$, where the wind intercepts the disk head-on ($\gamma \sim 90^{\circ}$) and the mass entrainment rate $\dot{\Sigma} \propto \sin^{2}\gamma$ could in principle be much higher.  However, the total mass entrainment rate in this case $\dot{M}_{\rm ws} \propto \dot{\Sigma}{\rm A} \propto {\rm A}c_{\rm s}$ must also be reduced due to the smaller surface area A of the rim relative to the disk annulus (by a factor $\sim H/R|_{R_{a}} \sim 0.1$) and the higher sound speed of the inner rim.  All things considered, the qualitative conclusion that $\tau_{\rm ws} \gtrsim \tau_{\rm life}$ for massive stars remains intact.  More detailed calculations, especially a more precise determination of the ``mixing efficiency'' $\xi$, will however be required to confirm this conclusion.

\subsection{Giant LBV Eruptions}
\label{sec:LBV}

A final means to disperse the disk is via singular, energetic eruptions from the star, such as those which may occur during the luminous blue variable (LBV) phase (e.g. \citealt{Conti84}; \citealt{Bohannan97}) of stars with masses $\gtrsim 40 M_{\sun}$ \citep{Langer+94}.  Assuming that the mass of the disk is concentrated near its outer edge, the gravitational binding energy of the disk is approximately
\begin{eqnarray}
&& E_{\rm bind} = \frac{GM_{\star}M_{d}}{2R}\approx 10^{47}{\,\rm ergs\,}M_{\star,100}^{2}(M_{d}/0.1M_{\star})R_{d,100}^{-1}.\nonumber \\
\end{eqnarray}
Because a fraction $\sim H/R|_{R_{d}} \sim 0.1$ of a spherically-symmetric outflow intercepts the disk (see Figs.~\ref{fig:predisk} and \ref{fig:diskfig}), an eruption with energy $\gtrsim 10^{48}$ ergs would in principle be sufficient to unbind a disk with mass $M_{d} \sim 10M_{\sun}$ from a 100 $M_{\sun}$ star.  In particular, $\eta-$Carinae's giant outburst in the 1840s \citep{Smith+03}, with an estimated energy $\sim 10^{50}$ ergs \citep{Smith08c}, could destroy even a massive proto-stellar disk.

Whether the disk in fact becomes unbound depends, however, on how efficiently the LBV outflow shares its kinetic energy with the disk.  To see why the disk-outflow coupling may in general be low, it is useful to draw an analogy with a similar physical system: a supernova shock impacting a molecular cloud (\citealt{McKee&Cowie75}; \citealt{McKee+78}; \citealt{Klein+94}).  When the LBV ejecta (with mass $M_{\rm ej}$ and expansion speed $v_{\rm ej}$) reaches the disk, a bow shock will form around the disk's inner edge.  The resulting pressure will drive a strong shock through the disk midplane at a velocity $v_{\rm sh} \sim v_{\rm ej}\chi^{-1/2}$ given by the shock jump conditions \citep{McKee&Cowie75}, where 
\be
\chi \equiv \frac{\rho_{d}}{\rho_{\rm ej}} \approx 7\left(\frac{H/R|_{R_{d}}}{0.1}\right)^{-1}\left(\frac{M_{d}}{M_{\rm ej}}\right)
\ee
is the ratio of the midplane density of the disk $\rho_{d} \approx M_{d}/2\pi R_{d}^{2}H|_{R_{d}}$ to the density of the LBV ejecta $\rho_{\rm ej} \approx M_{\rm ej}/(4\pi R_{d}^{3}/3)$.  

If the disk is more dense than the ejecta ($\chi > 1$), the time that the shock requires to cross the disk radially $t_{cr} \approx R_{d}/v_{\rm sh}$ (the ``crushing time,'' using the terminology of \citealt{Klein+94}) is a factor $\sim \chi^{1/2}$ longer than the expansion time the outflow takes to flow {\it around} the disk $t_{\rm exp} \sim R_{d}/v_{\rm ej}$.  Thus, when $\chi \gg 1$ the ram pressure of the outflow is not sustained sufficiently long to accelerate the entire disk to the post-shock speed $v_{\rm sh}$, much less to the ejecta speed (which requires several additional crushing times; \citealt{McKee+87}).  As a result, only a small portion of the outflow's kinetic energy is imparted to the disk.  A sufficiently dense disk should thus remain bound to the star, even if its binding energy is much less than the kinetic energy of the colliding ejecta.

During its giant eruption, $\eta-$Car ejected a mass $M_{\rm ej} \sim 10 M_{\sun}$ in less than a decade.  In this case $\chi \lesssim 1$ for $M_{d} \lesssim M_{\sun}$, implying that even a moderately massive disk could become unbound (see \citealt{Dwarkadas&Balick98}).   We note, however, that $\eta-$Car may represent an extreme case and its eruptions may not be representative of those from LBVs as a whole \citep{Vink09}.  Observations of the nebulae around LBVs typically indicate a wide distribution in the ejecta mass and energy ($M_{\rm ej} \sim 0.1-10M_{\sun}$; e.g. \citealt{Hutsemekers+94}; \citealt{Smith&Owocki06}), but in most cases it is unclear whether the ejecta primarily results from one or several giant eruptions \citep{Smith&Owocki06} or in a more continuous wind (e.g.~\citealt{Garcia-Segura+96}).  As discussed in $\S\ref{sec:wind}$, the latter case is unlikely to disperse the disk.  Even if LBV outflows are dominated by explosive events, we suspect that disks with masses $\gtrsim M_{\sun}$ may be sufficiently dense ($\chi \gg 1$) to remain bound, although hydrodynamical simulations will ultimately be required to precisely delineate how large $\chi$ must be for disk survival.

Incidentally, we note that were an LBV eruption to impact a relic proto-stellar disk, the resulting event (though not a supernova) would still be relatively luminous and could be contributing to the observed population of lower-luminosity transients known as ``supernova impostors'' \citep{VanDyk05}.

\subsection{Summary: Conditions for Disk Survival}
\label{sec:summary}

\begin{figure}
\resizebox{\hsize}{!}{\includegraphics[ ]{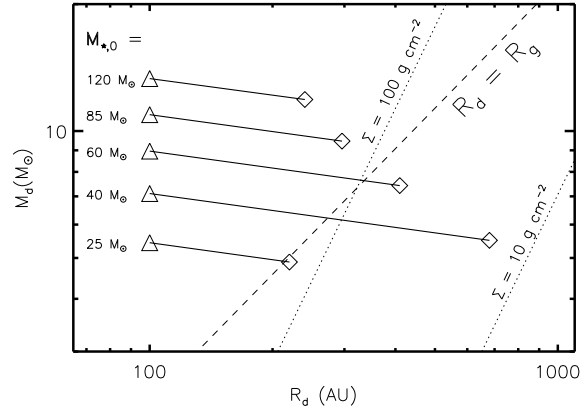}}
\caption{Time evolution of proto-stellar disks in the space of disk mass $M_{d}$ (left axis) and radius $R_{d}$ (bottom axis) for initial stellar masses in the range $M_{\star} = 25-120M_{\sun}$, calculated assuming that the disk resides in a state of marginal gravitational stability at $Q|_{R_{d}} = Q_{0} = 1$ ($\S\ref{sec:marginalGI}$) throughout the entire stellar lifetime, such that the disk spreads solely as the result of stellar mass loss.  Triangles show the initial location of the disk following the onset of its irradiation-supported phase, assuming that the initial radius of the disk is $\simeq$ 100 AU independent of stellar mass (KM06; see $\S\ref{sec:embed}$).  Diamonds show the final disk position at core collapse.  The point at which the radius of the disk equals the gravitational radius $R_{g}$ (eq.~[\ref{eq:rg}]) is shown with a dashed line.  Radii of constant surface density $\Sigma = M_{d}/A_{\rm gi}\pi R_{d}^{2}$ are shown with dotted lines for $\Sigma = 10$ and $\Sigma = 100$ g cm$^{-3}$.}
\label{fig:summary}
\end{figure}

Taken together, our results suggest that massive disks may preferentially survive until core collapse around {\it the most massive stars}.  One reason is that the most threatening dispersal processes, photo-evaporation ($\S\ref{sec:photo}$) and stellar collisions ($\S\ref{sec:collision}$), become increasingly effective for more radially-extended disks.  As shown in $\S\ref{sec:models}$, the disks around more massive stars tend to be more massive and compact at core collapse due to their higher initial masses; shorter stellar lifetimes; and less efficient angular momentum transport during the irradiation-supported phase, which results both from their higher surface densities and shielding from cosmic ray ionization by powerful stellar winds.  Everything else being equal, the more massive disks around massive stars are also less susceptible to destruction via stellar encounters and stellar wind stripping (simply because more mass requires longer to remove) and to giant LBV eruptions due to the inefficient coupling between a dense disk and lower-mass LBV outflows.  Important caveats to these conclusions include the large uncertainties in how the average cluster environment and the strength of LBV eruptions depend on stellar mass.  It is nevertheless plausible that the average disk mass at core collapse is a steeply increasing function of stellar mass.  This is consistent with the observation that long-lived massive disks are rare around B and late-type O stars  (\citealt{Natta+00}; \citealt{Fuente+06}), but suggests that disk survival should become more likely moving to early-type O stars.  

We emphasize that we {\it do not} propose that disk survival fraction approaches unity moving to the highest mass stars; rather, we are pointing to a trend within the confines of our assumptions.  There are still many ways that even disks around very massive stars may be destroyed, one of the most severe being the effect that a massive binary companion would have on the disk, a possibility which we have neglected from the onset, but which could nonetheless be a crucial factor in the majority of systems.  Indeed, if massive disks survived around all of even the highest mass stars (e.g. $M_{\star,0} \gtrsim 80 M_{\sun}$), the resulting number of VLSNe would be too high.

We encapsulate some of our conclusions in Figure \ref{fig:summary}, which shows the path taken by proto-stellar disks from the beginning of their isolated evolution until core collapse in the space of disk radius and disk mass for an initial radius $R_{d,0} = 100$ AU.  For concreteness we assume that angular momentum transport is negligible when the disk is irradiation-supported, such that the disks remains in a perpetual state of marginal gravitational instability ($\S\ref{sec:marginalGI}$) in which accretion and disk spreading occurs solely as the result of stellar mass loss.  For comparison we show the location where the disk radius equals the gravitational radius $R_{g}$ for photo-evaporation (eq.~[\ref{eq:rg}]) and contours of constant surface density $\Sigma = 10, 100$ g cm$^{-2}$.  Figure \ref{fig:summary} illustrates that disks around the highest massive stars spend the majority of their lives at radii $\ll R_{g}$ and with high surface densities ($\gg \Sigma_{a} \sim 10-100$ g cm$^{-2}$), i.e. those conditions which are most favorable for survival.

\section{Circumstellar Interaction in Core Collapse Supernovae}
\label{sec:diskshock}

Our results from Sections $\ref{sec:models}$ and $\ref{sec:destruct}$ suggest that when a proto-stellar disk survives until core collapse, its mass, outer radius, and scale height will take values in the range $M_{d} \sim 1-10M_{\sun}$, $R_{d} \sim$ few hundred AU, and $H/R|_{R_{d}} \sim 0.1-0.2$, respectively.  Once the expanding stellar ejecta from the supernova explosion impacts the disk, a portion of its kinetic energy will be thermalized, potentially resulting in bright optical or X-ray emission.  In this Section, we discuss the interaction between the disk and ejecta and its observable emission.

\begin{figure}
\resizebox{\hsize}{!}{\includegraphics[ ]{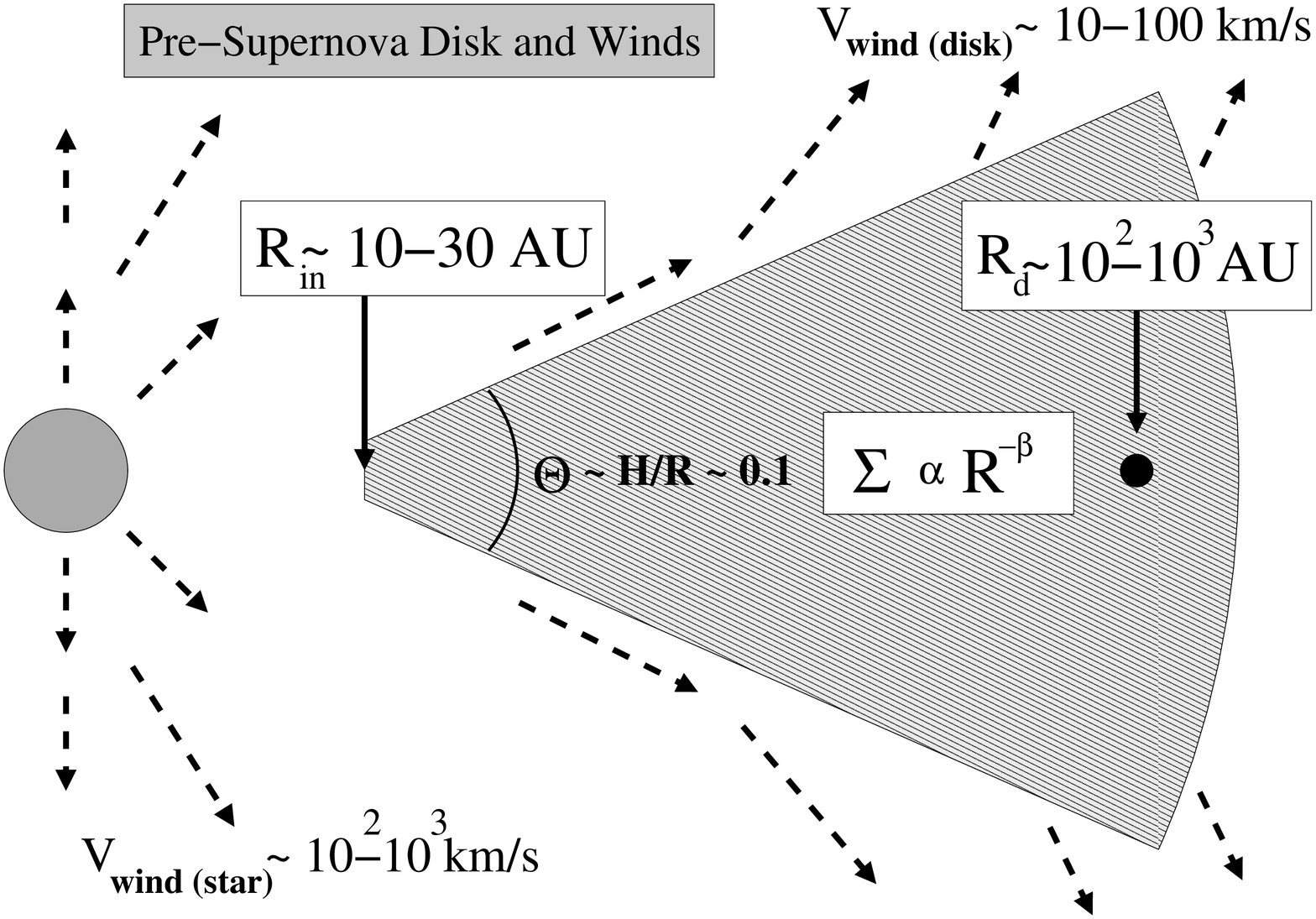}}
\caption{Pre-explosion properties of the relic star-disk system and associated outflows (see Table \ref{table:death}).  The fractional solid angle subtended by the disk is $H/R|_{R_{d}} \sim 0.1-0.2$.  The inner edge of the disk is located at $R_{a} \sim 10-30$ AU (eq.~[\ref{eq:rion}]) because interior to this radius the MRI is fully active and accretion is rapid.  The outer radius of the disk is $R_{d} \sim 10^{2}-10^{3}$ AU.  The surface density between $R_{\rm in}$ and $R_{d}$ obeys $\Sigma \propto R^{-\beta}$, where the value of $\beta \sim 1-2$ depends on the efficacy of angular momentum transport during the irradiation-supported phase ($\S\ref{sec:irrad}$).  Winds from the star are generally fast, with typical speeds $\sim 10^{2}-10^{3}$ km s$^{-1}$.  Winds from the disk, driven either by MHD processes at small radii or photo-evaporation coupled with radiation pressure on dust at larger radii, are generally slower, with speeds $\sim 10-100$ km s$^{-1}$.  Intermediate velocity outflows may originate in the mixing layer between the stellar wind and the disk surface \citep{Matsuyama+09}, or as the result of [shock-mediated] mixing between the stellar and disk winds.}
\label{fig:predisk}
\end{figure}

\begin{figure}
\resizebox{\hsize}{!}{\includegraphics[ ]{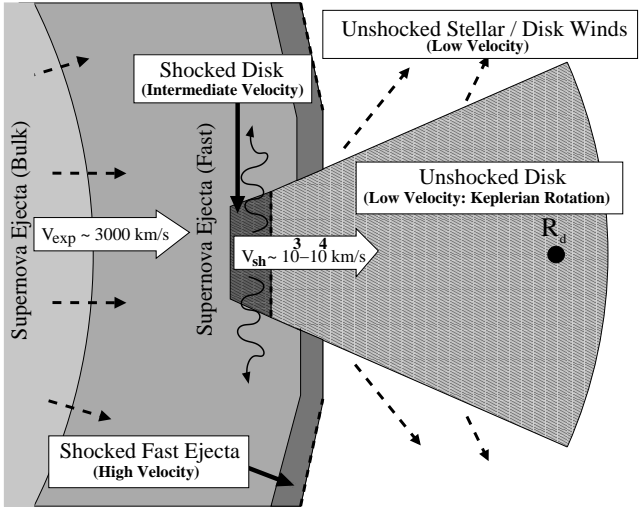}}
\caption{Interaction between the outgoing supernova ejecta and a relic disk.  The bulk of the SN mass and kinetic energy expands at a velocity $v_{\rm exp} \sim 3000$ km s$^{-1}$ (eq.~[\ref{eq:vexp}]).   The outer, fastest portions of the ejecta expand at the speed $v_{\rm fast} \sim 10^{4}$ km s$^{-1}$ and, upon reaching the inner edge of the disk at $R \sim R_{\rm in}$, drives a strong shock through the midplane.  The shock moves through the disk at a speed $v_{\rm sh}$ that varies from $v_{\rm sh} \sim v_{\rm fast} \sim 10^{4}$ km s$^{-1}$ at $R \sim R_{\rm in}$ to $v_{\rm sh} \sim 10^{3}$ km s$^{-1} < v_{\rm exp}$ at $R \sim R_{d}$, depending on the disk mass that has been accumulated.  If photons diffuse out of the midplane and cool the disk faster than the SN expansion timescale (viz.~$t_{\rm cool} \ll t_{\exp}$), the light curve directly traces the shock-generated thermal energy (eq.~[\ref{eq:edotsh}]; Fig.~\ref{fig:lc1}) and the SN spectrum may show evidence for CSM interaction, either through intermediate velocity lines ($\sim 10^{3}$ km s$^{-1}$) from the shocked disk or low velocity lines from the unshocked disk or stellar/disk winds.  If instead $t_{\rm cool} \gg t_{\rm exp}$, the energy generated by the shocked disk must diffuse through the envelope of the entire exploding star (Fig.~\ref{fig:lc2}), thereby displaying little direct evidence for CSM interaction.}
\label{fig:diskfig}
\end{figure}

\subsection{Ejecta-Disk Interaction}

We begin with a general discussion of the kinematics of the disk-ejecta collision.  Of the total kinetic energy $E_{\rm SN} \sim 10^{51}-10^{52}$ ergs and mass $M_{\rm ej} \approx M_{\star} \sim 10-100M_{\sun}$ of the SN ejecta\footnote{We assume that the SN explosion is largely successful in that most of the star is unbound, with only a modest portion of the final stellar mass used to form the neutron star or black hole compact remnant.}, only a fraction $f_{\Omega} \equiv \Omega_{d}/4\pi \approx H/R|_{R_{d}}$ will intercept the disk, where $\Omega_{d} \approx 4\pi(H/R)$ is the solid angle subtended by the disk from the star.  We consider SN energies up to $\sim 10^{52}$ ergs, above the canonical value of $10^{51}$ ergs, because VLSNe likely originate from very high mass stars (\citealt{Gal-Yam+07}; \citealt{Smith+07}) and their SNe often appear to be intrinsically ``hyper-''energetic (e.g.~\citealt{Rest+09}).

In the crude, yet instructive approximation that the interaction can be treated as a ``sticky'' collision between two masses $M_{1} = f_{\Omega}M_{\star}$ and $M_{2} = M_{d}$, the maximum efficiency for converting the SN kinetic energy into thermal energy $E_{\rm th}$ can be shown through conservation of energy and momentum to be
\be
\epsilon_{\rm max} \equiv \frac{E_{\rm th}|_{\rm max}}{E_{\rm SN}} = f_{\Omega}\frac{M_{2}}{M_{1} + M_{2}} = \left.\frac{H}{R}\right|_{R_{d}}\left(\frac{1}{1 + Q_{d}/A}\right),
\label{eq:erad}
\ee 
where $Q_{d} \gtrsim Q_{0} \sim 1$ is the Toomre parameter (eq.~[\ref{eq:ToomreQ}]) evaluated at $R = R_{d}$ and the constant $A = A_{\rm gi}-A_{\rm irr} \approx 2-6.8$ relates the total disk mass to the surface density at $R_{d}$ (see Appendix A).  

Equation (\ref{eq:erad}) shows that if the disk does not viscously evolve when irradiation-supported (such that $Q_{d} \sim Q_{0} \sim 1$ and $A = A_{\rm gi} \approx 2$ at core collapse; $\S\ref{sec:marginalGI}$), then $\epsilon_{\rm max} \sim H/R|_{R_{d}} \sim 10-20$ per cent.  Even if accretion is appreciable during the irradiation-supported phase, disks that preferentially survive have $Q_{d} \sim$ few $\lesssim A = A_{\rm irr} = 6.8$ at core collapse (see Fig.~\ref{fig:85msun}), again resulting in $\epsilon_{\rm max} \sim $ 10 per cent.  If the thermal energy generated by the collision is efficiently radiated, the total electromagnetic output may thus reach $\sim \epsilon_{\rm max}E_{\rm SN} \sim 10^{50}-10^{51}$ ergs, significantly higher than that from most core-collapse SNe and consistent with the optical energies of very luminous SNe.  We note that $\epsilon_{\rm max}$ and the corresponding luminosity could be even larger if the effective solid angle subtended by the disk is larger than our baseline estimate, due to e.g. a puffed up inner disk edge (e.g. \citealt{Dullemond+01}) or disk warping (e.g. \citealt{Pringle96}).  

We now discuss the radial structure of the disk prior to core collapse in order to explore the star-disk interaction in more detail (see Figure \ref{fig:predisk} for an illustration).  The inner edge of the disk occurs at $R_{\rm in} \approx R_{a} \sim 10-30$ AU (eq.~[\ref{eq:rion}]) because interior to this radius the MRI is fully active and accretion is rapid (see $\S\ref{sec:active}$).  Assuming that the disk has reached a steady state, the accretion rate $\dot{M} \propto \nu \Sigma$ between $R_{\rm in}$ and $R_{d}$ is approximately radially constant, where $\nu = \alpha c_{\rm s}^{2}/\Omega \propto \alpha TR^{3/2}$.  If the viscosity is very low in the irradiated state and the disk's structure at core collapse reflects that of marginal gravitational stability ($\S\ref{sec:marginalGI}$), then $\alpha = \alpha_{\rm gi} \propto \Sigma^{4}R^{21/2}$ (eq.~[\ref{eq:alpha_GI}]) and $T \propto \Sigma^{2} R^{3}$ (eq.~[\ref{eq:T_GI}]) and hence $\Sigma \propto R^{-15/7}$ (see Fig.~\ref{fig:diskevo} in Appendix A).  On the other hand, if the disk viscously evolves while irradiation-supported ($\S\ref{sec:alpha}$) then $T \propto R^{-3/8}$ (eq.~[\ref{eq:T_irrad}]) and hence $\Sigma \propto R^{-15/14}$ for constant $\alpha$.  In general we thus have $\Sigma \propto R^{-\beta}$, where $\beta \sim 1-2$.     

Properly normalized, the pre-supernova surface density thus takes the approximate form (for $R > R_{\rm in}$)
\begin{eqnarray}
\Sigma \simeq
10{\rm\, g\,cm^{-2}}\left(\frac{M_{d}}{M_{\sun}}\right)\left(\frac{R_{d}}{300{\,\rm AU}}\right)^{-2}\left(\frac{R}{R_{d}}\right)^{-\beta}\exp[-R/R_{d}], \nonumber \\
\label{eq:sigmapre} 
\end{eqnarray}
where in normalizing we have assumed that $R_{d} \sim 10^{2}R_{\rm in}$ and neglect the [weak] dependence of the pre-factor on $\beta \sim 1-2$.  We now scale the disk radius to a somewhat larger value $R_{d} = 300$ AU, which is more typical of the final radius at core collapse (Table \ref{table:death}).  Exterior to $R_{\rm d}$ we have assumed that $\Sigma$ decreases exponentially with radius, as would be expected from the isolated viscous evolution of an $\alpha-$disk (e.g. \citealt{Pringle81}; see Appendix A).  In reality, the surface density outside $R_{d}$ may decrease more rapidly with radius if, for instance, the disk is marginally gravitationally stable ($\S\ref{sec:marginalGI}$; see Fig.~\ref{fig:diskevo}) or if the outer edge has been truncated by photo-evaporation ($\S\ref{sec:photo}$) or a recent stellar collision ($\S\ref{sec:collision}$). 

The stellar ejecta expands at an average velocity 
\begin{eqnarray}
&&v_{\rm exp} = \left(\frac{2E_{\rm SN}}{M_{\rm ej}}\right)^{1/2} \nonumber \\
&&\approx 3\times 10^{3}{\rm km\,s^{-1}\,}\left(\frac{E_{\rm SN}}{10^{52}{\,\rm ergs}}\right)^{1/2}\left(\frac{M_{\rm ej}}{100M_{\sun}}\right)^{-1/2}
\label{eq:vexp}
\end{eqnarray}
and reaches the inner edge of the disk on a timescale $R_{\rm in}/v_{\rm exp} \sim $ days to a week.  Although $v_{\rm exp}$ is the average ejecta speed, the expansion is homologous with a broad velocity distribution (e.g.~\citealt{Chevalier&Fransson94}).  In particular, the outer ejecta expands at a typical rate $v_{\rm fast} \sim 10^{4}$ km s$^{-1}$ which can be several times larger than $v_{\rm exp}$.  Once this fast ejecta reaches the inner edge of the disk, a bow shock will form and a strong shock will be driven through the midplane at the speed $v_{\rm sh}$ (similar to the interaction with giant LBV eruptions; $\S\ref{sec:LBV}$).  This situation is illustrated in Figure \ref{fig:diskfig}.  The shock through the disk generates thermal energy at the rate
\begin{eqnarray}
\dot{E}_{\rm sh}&& \simeq (1/2)\psi\rho_{d}v_{\rm sh}^{3}\times(4\pi f_{\Omega}R^{2}) \simeq \pi\psi\Sigma R v_{\rm s}^{3} \nonumber \\
&& \approx 1.4\times 10^{44}{\,\rm ergs\,s^{-1}}\psi\left(\frac{v_{\rm sh}}{10^{4}{\,\rm km\,s^{-1}}}\right)^{3}\times \nonumber \\
&& \left(\frac{M_{d}}{M_{\sun}}\right)\left(\frac{R_{d}}{300{\,\rm AU}}\right)^{-1}\left(\frac{R}{R_{d}}\right)^{1-\beta}\exp[-R/R_{d}],
\label{eq:edotsh}
\end{eqnarray}
where $\rho_{d} \simeq \Sigma/2H$ is the midplane density, $\psi \approx 1$ is an efficiency factor, and we have used equation (\ref{eq:sigmapre}) for $\Sigma$.    

The ratio of the disk density to the mean ejecta density $\bar{\rho}_{\rm ej} \equiv M_{\star}/(4\pi R^{3}/3)$ at radius $R$ can be written as
\be
\chi\equiv \frac{\rho_{d}}{\bar{\rho}_{\rm ej}} \approx 1.8 Q_{d}^{-1}(R/R_{\rm d})^{(12/7)-\beta}\exp[-R/R_{d}],
\ee
where we have used equation (\ref{eq:hoverr_irrad}) for $H/R$.  Note that $\chi \lesssim 1$ is satisfied at all radii for all physical disks (i.e. those with $Q_{d} \gtrsim 1$).  Thus, unlike the case of giant LBV eruptions described in $\S\ref{sec:LBV}$, the efficiency with which the SN ejecta shares its energy with the disk material is necessarily high. 

If viscous evolution occurs during the irradiation-supported phase (such that $\beta = 15/14$ and $Q_{d} \sim$ few) the disk mass $\propto \Sigma R^{2} \propto R^{13/14}$ is concentrated at large radii and $\chi$ remains $\ll 1$ for $R \ll R_{d}$.  The stellar ejecta and the swept-up disk mass thus only comparable (and hence deceleration becomes significant) once the shock reaches radii $\sim R_{d}$.  If, on the other hand, the disk is marginally gravitationally stable at collapse (such that $\beta \sim 27/14$ and $Q_{d} \sim Q_{0} \sim 1$; $\S\ref{sec:marginalGI}$) appreciable deceleration thus begins almost immediately at $R \gtrsim R_{\rm in}$ because the disk mass is spread out approximately equally per radial decade (see Fig.~\ref{fig:diskevo}).

\subsection{Two Regimes of Interaction}
\label{sec:tworegimes}

In this section we address the emission produced by the disk-star interaction.  The models we present are not intended to describe any supernova in particular and are less detailed than other models of CSM interaction available in the literature (e.g.~\citealt{Chugai+04}; \citealt{Chugai&Chevalier06}).  Rather, our goal is to qualitatively illustrate the diversity of observable light curve and spectral behavior that may manifest from the ejecta-disk interaction and how this may relate to observed features of luminous SNe.  More detailed radiation hydrodynamic simulations will ultimately be required to confirm and refine the ideas presented here.  Although in general the ejecta will possess additional sources of energy, due to e.g.~radioactive decay or residual thermal energy, we focus on the luminosity that results solely from the interaction with the disk.  

Although $\dot{E}_{\rm sh}$ (eq.~[\ref{eq:edotsh}]) represents the power generated as the shock passes through the disk, this thermal energy may not be immediately radiated.  If the disk's opacity is $\kappa$, the vertical optical depth is $\tau_{d} \approx \Sigma\kappa/2$ and the photon diffusion time out of the midplane (for $R_{\rm in} \lesssim R \lesssim R_{d}$) is given by:
\begin{eqnarray}
&&t_{\rm diff,d} = t_{\rm cool} \approx (H/c)\tau_{d} \approx3{\rm\,days\,}\left(\frac{M_{d}}{M_{\sun}}\right)\times \nonumber \\
&&   \left(\frac{H/R}{0.1}\right)\left(\frac{R_{d}}{300{\,\rm AU}}\right)^{-1}\left(\frac{R}{R_{d}}\right)^{1-\beta}\left(\frac{\kappa}{\kappa_{es}}\right), \nonumber \\
\end{eqnarray}
where we have scaled $\kappa$ to the electron scattering opacity $\kappa_{es} \approx 0.4$ cm$^{2}$ g$^{-1}$, a reasonable approximation for the very high temperatures $T \gtrsim 10^{7-8}$ K behind the shock.  For $\Sigma$ we use the post-shock surface density, which is larger than the pre-shock value (eq.~[\ref{eq:sigmapre}]) by a factor $\simeq 7$, which represents the radial compression due to a strong, adiabatic, radiation pressure-dominated shock.\footnote{Note that if $t_{\rm cool}$ is sufficiently short, the shock will be radiative and the post-shock density may become significantly higher than just due to adiabatic compression.  The adiabatic approximation is nevertheless valid on timescales $\lesssim t_{\rm cool}$, thereby justifying its use in equation (\ref{eq:diffexpratio}).}  Since the post-shock pressure is dominated by radiation, $t_{\rm diff,d}$ is also the cooling timescale of the disk $t_{\rm cool}$ (eq.~[\ref{eq:tcool}]).  

As we show below, the properties of the emission from the shocked disk at radius $R$ depends on the ratio of the cooling timescale $t_{\rm cool}$ to the average expansion timescale of the bulk of the ejecta $t_{\rm exp} \simeq R/v_{\rm exp}$:
\begin{eqnarray}
\frac{t_{\rm cool}}{t_{\rm exp}}&& \simeq 0.2 \left(\frac{H/R}{0.1}\right)\left(\frac{E_{\rm SN}}{10^{52}{\,\rm ergs}}\right)^{1/2}\left(\frac{M_{\star}}{100M_{\sun}}\right)^{1/2}\times \nonumber \\
 &&\left(\frac{M_{d}}{0.1M_{\star}}\right)\left(\frac{R_{d}}{300{\,\rm AU}}\right)^{-2}\left(\frac{R}{R_{d}}\right)^{-\beta},
\label{eq:diffexpratio}
\end{eqnarray}
where we have used equation (\ref{eq:vexp}) for $v_{\rm exp}$ and have assumed $M_{\rm ej} \approx M_{\star}$.

\subsubsection{Fast Cooling Case}
\label{sec:fastcool}

Low mass and/or radially-extended disks satisfy the condition $t_{\rm cool} \ll t_{\rm exp}$.  In this case the thermal energy generated by the shock escapes the disk before it is engulfed by the expanding optically-thick star (see Fig.~\ref{fig:diskfig}).  The shock will thus be radiative and the swept-up disk will cool into a thin shell (e.g.~\citealt{Chugai01}).  X-rays generated behind the shock may escape to infinity, producing detectable high-energy emission (e.g. \citealt{Immler&Kuntz05}; \citealt{Schlegel&Petre06}), or they may be absorbed and reprocessed into optical (e.g. H$\alpha$) emission (e.g. \citealt{Fransson+96}).  The observed bolometric light curve thus directly tracks the shock-generated power, i.e. $L_{\rm bol} \approx \dot{E}_{\rm sh}$. 

Since the swept up disk mass becomes comparable to the ejecta mass at radii $R \lesssim R_{d}$, the cool shell will generally decelerate below the expansion speed of the blast wave.  Photons re-radiated by the disk are thus a possible origin for the  ``intermediate velocity'' lines with typical widths of a few thousand km s$^{-1}$, as are sometimes observed in Type IIn SNe (e.g. \citealt{Chugai&Danziger94}).  
\begin{figure}
\resizebox{\hsize}{!}{\includegraphics[ ]{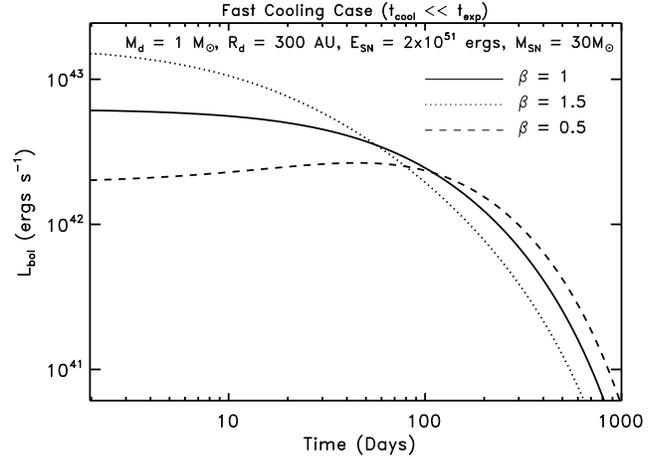}}
\caption{Light curves from the shock-heated disk in the fast cooling case $t_{\rm cool} \ll t_{\rm exp}$ ($\S\ref{sec:fastcool}$), calculated for a total stellar ejecta mass $M_{\star} = 30 M_{\sun}$; supernova energy $E_{\rm SN} = 2\times 10^{51}$ ergs; and disk mass $M_{d} = 1M_{\sun}$ and outer radius $R_{d,f} = 300$ AU, respectively.  We show calculations for different values of the surface density index $\beta$, where $\Sigma \propto R^{-\beta}$ (eq.~[\ref{eq:sigmapre}]).  }
\label{fig:lc1}
\end{figure}

Figure \ref{fig:lc1} shows the bolometric light curve $L_{\rm bol} = \dot{E}_{\rm sh}$ in the fast-cooling case, calculated from equation (\ref{eq:edotsh}) for a disk with mass $M_{d} = 1M_{\sun}$ and radius $R_{d} = 300$ AU; supernova energy $E_{\rm SN} = 2\times 10^{51}$ ergs and ejecta mass $M_{\rm ej} = 30 M_{\sun}$; and for different values of surface density profile index $\beta$.  The shock heating rate $\dot{E}_{\rm sh}$ is calculated using equation (\ref{eq:edotsh}), accounting for the deceleration of the shock speed $v_{\rm sh}(t)$ using the simple thin shell model given in \citet{Chugai&Danziger94} in which the total momentum $M_{\rm ej}f_{\Omega}v_{\rm exp}$ is conserved.  We take $f_{\Omega} = H/R = 0.1$ as the effective scaleheight in this one-dimensional model.    

For all values of $\beta$ the total radiated energy is $E_{\rm bol} \approx 7\times 10^{49}$ ergs or $\sim 5$ per cent of the SN kinetic energy, significantly larger than the electromagnetic output of most core-collapse SNe (e.g.~\citealt{Bersten&Hamuy09}).  Also note that for $\beta = 0.5$, the emission peaks at late times, only once the shock reaches the outer edge of the disk at $t_{\rm peak} \approx 100$ days.  Such behavior provides one explanation for events with long rise times until peak emission (e.g.~SN 2008iy; \citealt{Miller+09}).

\subsubsection{Slow Cooling Case}
\label{sec:slowcool}

More massive, compact disks instead have $t_{\rm cool} \gg t_{\rm exp}$.  In this case the shocked thermal energy remains trapped until after the disk is engulfed by the expanding, optically-thick star.  As a result, the shock-generated power $\dot{E}_{\rm sh}$ must instead diffuse out of the {\it entire} stellar envelope, which occurs on a longer envelope diffusion timescale
\be
t_{\rm diff,\star} = \frac{B\kappa M_{\star}}{c R},
\ee
where $\kappa$ is the opacity and $B \simeq 0.07$ for a spherical outflow (e.g. \citealt{Padmanabhan00}).  

\begin{figure}
\resizebox{\hsize}{!}{\includegraphics[ ]{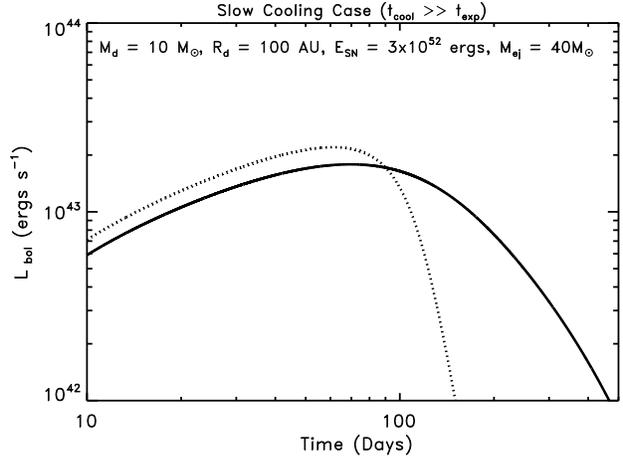}}
\caption{Light curve powered by the shocked disk in the slow cooling case $t_{\rm cool} \ll t_{\rm exp}$ ($\S\ref{sec:slowcool}$), calculated for a total stellar ejecta mass $M_{\star} = 40 M_{\sun}$; supernova energy $E_{\rm SN} = 3\times 10^{52}$ ergs; and disk mass $M_{d} = 10M_{\sun}$ and outer radius $R_{d} = 100$ AU, respectively.  We take $\beta = 1$ in the disk density profile.  The model shown with a solid line assumes that the disk surface density outside $R_{d}$ decreases exponentially with radius, as expected from the isolated viscous evolution of an irradiation-supported, constant-$\alpha$ disk (eq.~[\ref{eq:sigmapre}]).  The calculation shown with a dashed line instead assumes a sharper outer disk edge, as would be expected if the disk resides in a state of marginal gravitational instability at collapse ($\S\ref{sec:marginalGI}$; see Fig.~\ref{fig:diskevo}) or if the outer edge has been truncated due to, e.g., photo-evaporation ($\S\ref{sec:photo}$) or a recent stellar collision ($\S\ref{sec:collision}$).}
\label{fig:lc2}
\end{figure}

Figure \ref{fig:lc2} shows an example model for the bolometric light curve in the slow cooling case, calculated for a disk with mass $M_{d} = 10 M_{\sun}$ and radius $R_{d} = 200$ AU; supernova energy $E_{\rm SN} = 3\times 10^{52}$ ergs; ejecta mass $M_{\rm ej} = 40M_{\sun}$.  We assume the disk density index is $\beta=1$ and the scaleheight $H/R = 0.1$.  We calculate the light curve using a simple one-zone diffusion model, in which the radiated luminosity is given by $L = E_{\rm th}/t_{\rm diff,\star}$, where $E_{\rm th}$ is the thermal energy of the ejecta (e.g. \citealt{Li&Paczynski98}).  We evolve $E_{\rm th}$ in time as the star expands, including adiabatic losses due to PdV work and time-dependent heating from the shock-generated energy $\dot{E}_{\rm sh}$.  As in the fast-cooling case, we calculate $v_{\rm sh}(t)$ using a momentum-conserving thin-shell model.  We assume that the opacity is electron scattering $\kappa \approx \kappa_{es} \sim 0.4$ cm$^{2}$ g$^{-1}$.   

The two cases shown in Figure \ref{fig:lc2} correspond to different assumptions regarding the density profile at the outer edge of the disk.  In the model shown with a solid line, we have assumed that the surface density decreases exponentially with radius $R_{d}$ (eq.~[\ref{eq:sigmapre}]), as expected from the isolated viscous evolution of an irradiation-supported constant-$\alpha$ disk.  By contrast, the dotted line shows a model for a sharper outer disk edge, as appropriate for a disk which is marginally gravitationally-stable ($\S\ref{sec:marginalGI}$; Fig.~\ref{fig:diskevo}) or has its outer edge truncated due to disk winds ($\S\ref{sec:photo}$; $\S\ref{sec:wind}$) or stellar collisions ($\S\ref{sec:collision}$).  In both cases we find that the total radiated energy is $\sim 3\times 10^{50}$ ergs.  The radiative efficiency is lower than in the fast cooling case because a portion of the trapped thermal energy is lost to PdV work before diffusing out as radiation.  Because the ratio $t_{\rm exp}/t_{\rm diff,\star}$ increases rapidly with radius ($\propto R^{2}$), adiabatic losses are, however, mild compared to those incurred from the compact stellar surface.  

Although we have discussed the fast and slow cooling cases as if they are mutually exclusive, in general $t_{\rm cool}/t_{\rm exp}$ will vary with radius and the resulting emission will be more complex than the two limiting cases described above.  For instance, $t_{\rm cool}/t_{\rm exp} \gtrsim 1$ may be satisfied at small radii, in which case a portion of the thermal energy generated by the shocked inner disk will be trapped behind the stellar photosphere.  However, if $t_{\rm cool}/t_{\rm exp} \propto R^{-\beta} \lesssim 1$ is satisfied once as the shock reaches the outer disk, a significant portion of the thermal energy could also be radiated promptly.  This type of behavior could explain ``transition'' events like SN 2006tf, which \citet{Smith+08} argue evolves from an optically-thick, ``diffusive'' phase to emission which shows more direct evidence for CSM interaction at later times.

\section{Discussion}
\label{sec:discussion}

\subsection{Long-Lived Disks around Very Massive Stars}
\label{sec:disks}

A straightforward prediction of the relic disk model is that a small fraction of very massive stars should possess a proto-stellar disk at radii of tens to hundreds of AU {\it throughout their entire lifetime} (including putative post-main sequence LBV and Wolf-Rayet phases).  The chief difficulties in testing this prediction are that (1) massive stars are typically far away, and their environments tend to be messy, crowded, and obscured (\citealt{Cesaroni+07}; \citealt{Zinnecker&Yorke07}); and (2) because very luminous SNe are very rare, only a small fraction of massive stars must retain a disk in order to account for VLSNe via CSM interaction with a relic disk.  For instance, if a fraction $f\sim 10^{-2}-10^{-4}$ of all core collapse events produce VLSNe, then at any time we would expect to find $N \sim fR_{\rm SN}t_{\rm life} \sim 3-300$ progenitors in a galaxy like the Milky Way, where $R_{\rm SN} \sim 10^{-2}$ yr$^{-1}$ is the total core collapse rate and $t_{\rm life} \sim 3$ Myr.  We now discuss whether such progenitor systems actually exist.

It is clear that long-lived disks are not present around the majority of [what are normally considered] main sequence O stars.  The emission line profiles of some main sequence O stars (the ``Oe'' and ``Onfp'' classes) have long been interpreted as indicating the presence of an ionized disk \citep{Conti&Leep74}.  Spectropolarimetry of O stars (\citealt{Harries+02}; \citealt{Vink+09}), LBVs (e.g.~\citealt{Schulte-Ladbeck+93}), and Wolf-Rayet stars (e.g.~\citealt{Vink07}) also suggest that the environments of massive stars are asymmetric in some cases.  These probes are, however, primarily sensitive to scales of only a few stellar radii, and the observed polarization may be caused by intrinsic asymmetries in the stellar wind, perhaps due to rapid rotation \citep{Bjorkman&Cassinelli93}.  Furthermore, massive irradiated disks around otherwise unobscured stars would likely produce extremely bright emission line (e.g. H$\alpha$) luminosities, which are not observed.

A long-lived disk should also be a strong source of infrared (IR) emission.  Since the disk scaleheight is $H/R \sim 0.1$, if the disk is flared then at least $\sim 10$ per cent of the stellar luminosity will be absorbed and re-radiated as thermal blackbody and dust emission peaked at near- to mid-IR wavelengths (e.g.~\citealt{Chiang&Goldreich97}; \citealt{Rafikov&DeColle06}).  Excess IR emission is in fact common from very massive stars during essentially all stages of evolution.  For instance, \citet{Barniske+08} find evidence for mid-IR dust emission down to scales $\lesssim 100$ AU around two nitrogen-rich WN stars near the Galactic center, which they infer to be among the most luminous stars in the Milky Way (see also~\citealt{Rajagopal+09}).  Wolf-Rayet (WR) stars, especially of the carbon sequence (WC), commonly radiate a significant fraction ($\sim 0.1-10\%$) of their total luminosity in the IR (e.g.~\citealt{Williams+87}; \citealt{Hadfield+07}).  In cases when the infrared emission is partially resolved, the inferred radii are $\sim 30-300$ AU (\citealt{Dyck+84}; \citealt{Ragland&Richichi99}), consistent with the expected scale of a relic disk.  Excess IR emission is also common from LBVs and, when marginally resolved, size scales of tens of AU are commonly inferred (\citealt{Rajagopal+07}).  Although these hints are suggestive, the mid-IR emission from massive stars is generally well-modeled as originating from a quasi-spherical, dusty outflow, without the need to invoke a disk (e.g.~\citealt{Williams+87}; \citealt{Crowther03}).

Although there appears to be no evidence for long-lived disks around unobscured stars, it is important to keep in mind that the time-sequence and durations of the observable phases of massive stellar evolution remain uncertain \citep{Zinnecker&Yorke07}.  When massive stars are very young and still actively accreting, they may be observed as ``hot molecular cores,'' with large masses of warm and dense gas (\citealt{Kurtz+00}; \citealt{Cesaroni05}).  Once external accretion shuts off, however, the system is thought to transition to a compact HII phase (\citealt{Kurtz05}; \citealt{Hoare+07}), during which the central star begins ionizing pockets of surrounding gas.  

After the HII phase, the system is generally thought to emerge as an unobscured main sequence star.  For massive stars, however, the HII phase is known to last a significant fraction ($\sim 10$ per cent, on average) of the stellar lifetime \citep{Wood&Churchwell89}.  In order to explain the anomalously long lifetimes of compact HII regions, \citet{Hollenbach+94} proposed that they are continually replenished with material from a photo-evaporating disk; this model in fact remains a leading explanation for {\it hyper-compact} HII regions (\citealt{Keto07}; \citealt{Nielbock+07}) with electron number densities $n_{e} \gtrsim 10^{5-6}$ cm$^{-3}$ and inferred sizes $\lesssim 10^{3}$ AU (e.g.~\citealt{Kurtz+00}; \citealt{DePree+05}).  Although disks around the most massive stars may not completely photo-evaporate ($\S\ref{sec:photo}$) or be stripped by stellar winds ($\S\ref{sec:wind}$), the mass loss rate from the disk can still be quite high, such that several solar masses could be lost from the disk throughout the stellar lifetime.  By producing a dusty shell around the star, a continuous disk outflow could in principle cause even a $\sim$ Myr old star to appear to be in a younger stage of evolution.

To summarize, if VLSNe indeed originate from relic proto-stellar disks then their progenitors should manifest as conspicuous sources of IR emission and possibly as long-lived hyper-compact HII regions.  If unobscured, such systems should be exceedingly bright sources of optical (e.g.~H$\alpha$) line emission.  Although no such progenitor population has yet been identified, only a small fraction of massive stars produce VLSNe and evolved stars with relic disk systems could in principle be confused with younger embedded systems.  We note that only quite recently are blind samples of hyper-compact HII regions \citep{Murphy+10}, and IR-selected samples of evolved massive stars (\citealt{Hadfield+07}; \citealt{Shara+09}; \citealt{Carey+09}) and the general population of self-obscured massive stars (\citealt{Thompson+09}; \citealt{Khan+10}), being assembled.  The possibility that a population of relic disk systems may exist without our knowledge is underscored by the recent discovery of a ``hidden'' population of massive stars with the {\it Spitzer Space Telescope} \citep{Wachter+10}.

\subsection{Very Luminous Supernovae}
\label{sec:type2n}

Many VLSNe are classified as Type IIn (\citealt{Schlegel90}; \citealt{Filippenko97}), which are characterized by (1) slowly-decaying light curves that can persist for decades (e.g.~\citealt{Miller+10}); (2) very narrow (widths of tens to a few hundred km s$^{-1}$) H emission lines, which are thought to arise from photo-ionized unshocked CSM; (3) intermediate-width (typically $1-5\times 10^{3}$ km s$^{-1}$) H lines, thought to arise from shock-accelerated CSM (e.g.~\citealt{Chugai&Danziger94}), (4) and, sometimes, luminous X-ray and radio emission for years after the explosion (e.g.~\citealt{Immler&Kuntz05}).  These characteristics are in addition to the broad lines with widths $\sim 10^{4}$ km s$^{-1}$ which are observed in most SNe.  The progenitors of IIn SNe are probably among the most massive stars (e.g.~\citealt{Gal-Yam&Leonard09}).

As described in $\S\ref{sec:diskshock}$, many of the properties of the most luminous SN IIn can be produced via the interaction of the supernova shock with a massive, relic proto-stellar disk (see Figure \ref{fig:diskfig}).  Broad lines may originate from the shocked fast supernova ejecta, which occupies the majority ($\sim 80-90\%$) of the solid angle around the exploding star; they may be powered by either the fast ejecta interacting directly with the dense wind from the star or disk wind; or through the absorption of X-rays radiated at mid latitudes as the shock passes through the disk.  Intermediate lines may originate from the shocked disk, which is initially accelerated by the fast ejecta, but decelerates once it has swept up appreciable mass.  Since the mass of the disk and the sector of the star which intercepts the disk are approximately equal (see eq.~[\ref{eq:erad}]), intermediate lines with widths $\sim 1/2$ of the bulk ejecta speed $v_{\rm exp}$ (eq.~[\ref{eq:vexp}]) around the time of peak emission are thus a natural prediction of the relic disk model.  

The narrow line emission in SNe IIn is generally well-modeled as originating from the outer, unshocked portion of the same outflow which is responsible for the bulk of the luminosity (e.g.~\citealt{Chugai+04}).  However, in the relic disk model, narrow line emission must originate from a region distinct from the disk itself.  Disk outflow due to photo-evaporation ($\S\ref{sec:photo}$) or stellar wind stripping ($\S\ref{sec:wind}$) are generally too slow to explain the line widths of up to several hundred km s$^{-1}$ that are observed in luminous SN IIn.  However, in principle the faster stellar wind and the slower disk material could mix, resulting in an outflow with intermediate velocities.  Indeed, {\it empirically} broad radio recombination lines with widths $\sim 30-200$ km s$^{-1}$ are observed in hyper-compact HII regions (\citealt{Gaume+95}; \citealt{Shepherd+95}; \citealt{Sewilo+04}), which are thought to result from photo-evaporating disks (\citealt{Hollenbach+94}; \citealt{Keto07}).

The majority of SN IIn probably cannot result from relic disks.  Since SN IIn represent up to $\sim 10$ per cent of core collapse SNe, the implied population of progenitors would probably have been identified ($\S\ref{sec:disks}$).  On the other hand, SN IIn are an inhomogeneous population and the origin of the CSM in the general population may be different than in the most luminous cases.  

We also emphasize that not all VLSNe are Type IIn.  Many of the most energetic events such as 2005ap \citep{Quimby+07} and 2008es (\citealt{Miller+09}; \citealt{Gezari+09}) lack intermediate width lines ($\sim 2000$ km s$^{-1}$), while the high-redshift PTF events show no evidence for H or He at all \citep{Quimby+09}.  Although on the surface these observations appear difficult to explain via interaction with a hydrogen-rich proto-stellar disk, evidence for CSM interaction could be masked if the disk is particularly massive and/or compact because it may not have time to radiate intermediate-width lines before being engulfed by the expanding ejecta (see $\S\ref{sec:slowcool}$ and Fig.~\ref{fig:lc2}).  Indeed, our calculations in $\S\ref{sec:models}$ suggest that massive, compact disks which satisfy $t_{\rm cool} \gg t_{\rm exp}$ (eq.~[\ref{eq:diffexpratio}]) are only expected to survive around the highest mass (and, hence, rarest) stars which may have the most energetic explosions.  

\subsection{``Hybrid'' Type I/IIn SNe}
\label{sec:hybrid}

SN 2002ic \citep{Hamuy+03} was classified as a peculiar Type Ia SN, but with the unusual feature of narrow and broad components of H$\alpha$ emission, suggesting the presence of strong H-rich CSM interaction as in SNe IIn.  This ``hybrid'' Type Ia/IIn SN was interpreted by \citet{Hamuy+03} as resulting from the thermonuclear disruption of a white dwarf into a hydrogen-rich environment, perhaps resulting from a pre-explosion AGB wind \citep{Hamuy+03} or common envelope ejection \citep{Livio&Riess03}.  The required CSM mass to explain the observed light curve was up to several solar masses on a radial scale $\sim 500$ AU \citep{Chugai&Yungelson04}.  SN 2005gj (\citealt{Prieto+05}; \citealt{Aldering+06}) also showed hybrid Type Ia/IIn characteristics similar to 2002ic, which again appeared to require a significant quantity of mass on a radial scale of a few hundred AU.

\citet{Benetti+06} recently questioned the identification of 2002ic as an intrinsic Type Ia SN, suggesting it should instead be classified Type Ic, events which are thought to result from the core collapse of WR stars.  Although it is unclear why a star would lose its entire hydrogen envelope just years prior to core collapse, hydrogen-rich material around an evolved star at a few hundred AU could be explained as a relic proto-stellar disk.  Spectropolarimetry of 2002ic supports this hypothesis by showing that the geometry was likely aspherical and, possibly, equatorially-concentrated (\citealt{Kotak+04}; \citealt{Wang+04}).  Analysis of the light curve (\citealt{Wood-Vasey+04}; \citealt{Uenishi+04}) further showed a delay between the time of explosion and the onset of the CSM interaction, suggesting the presence of a ``cavity'' between the star and the CSM.  A lower-density region in proto-stellar disks is expected interior to the MRI-active radius $R_{a} \sim 30$ AU (eq.~[\ref{eq:rion}]; see Fig.~\ref{fig:predisk}).    

\section{Conclusions}
\label{sec:conclusions}

Circumstellar interaction is a promising mechanism for powering some of the most luminous and energetic SNe yet discovered.  However, if the requisite CSM is the result of a pre-SN stellar eruption, up to several solar masses of material must be ejected just years$-$decades prior to explosion.  Although in principle giant LBV-like eruptions or pair instabilities are capable of producing such prodigious mass loss, there is currently no explanation for why the eruption and the stellar explosion are so nearly synchronized.
  
In this paper we have developed an alternative hypothesis that the bulk of the CSM is not an outflow (thus avoiding the ``coincidence'' problem), but rather a pre-existing, long-lived disk left over from stellar birth.  One virtue of our model is that a CSM mass up to $\sim 10 M_{\sun}$ is explained as the maximum disk mass that can be supported against self-gravity by stellar irradiation (eq.~[\ref{eq:diskmassratio_irr}]).  The CSM radii $\sim 10^{2}-10^{3}$ AU required by observations are also explained naturally by (1) the maximum outer disk radius ($\sim 100-200$ AU) that is stable to fragmentation during the embedded phase (\citealt{Kratter&Matzner06}; $\S\ref{sec:embed}$); and (2) the limited subsequent viscous evolution experienced by disks around massive stars ($\S\ref{sec:irrad}$; $\S\ref{sec:models}$).  Indeed, we find that massive disks may preferentially survive until core collapse around the most massive stars ($\S\ref{sec:summary}$) for the following combination of reasons (see also Fig.~\ref{fig:summary}):
\begin{itemize}
\item{Massive stars live short lives, thereby allowing less time for accretion and other disk dispersal processes to act.}
\item{Massive stars begin their irradiation-supported phase of evolution with more massive disks (eq.~[\ref{eq:diskmassratio_irr}]), which generally take longer to disperse.}
\item{Disks around massive stars have larger surface densities at the onset of irradiation-supported accretion (eq.~[\ref{eq:sigma0}]).  This shields the midplane from external sources of ionization, resulting in extensive and sustained dead-zones in which the magneto-rotational instability is quenched and angular momentum transport is inefficient.  The powerful stellar winds from massive stars appear likely to exclude interstellar cosmic rays, leaving X-rays (which penetrate to a shallower depth) as the chief source of surface ionization (eq.~[\ref{eq:rloverr}]).}
\item{Due to their slower expected viscous spreading and deeper gravitational potential, disks around massive stars are less likely to be dispersed via photo-evaporation ($\S\ref{sec:photo}$). }
\item{Massive disks can be denser than the ejecta from giant LBV eruptions, suggesting they are more likely to remain bound to the star (\S\ref{sec:LBV}).}
\end{itemize}
Our conclusion that the probability of disk survival rapidly increases with stellar mass is consistent with evidence that very luminous IIn SNe originate from the most massive stars (\citealt{Gal-Yam+07}; \citealt{Smith+07}).  We do not, however, suggest that the disk survival fraction approaches unity in any stellar mass range, as external influences, such as the presence of a massive binary companion or a particularly dense stellar environment, may preclude disk survival in the majority of cases.

Several of the observed properties of VLSNe can be explained by the collision between the supernova ejecta and a relic disk (see Figs.~\ref{fig:predisk} and \ref{fig:diskfig}).  The nominal radiative efficiency of the disk-ejecta interaction is $\sim 10$ per cent (eq.~[\ref{eq:erad}]), resulting in a total electromagnetic output $\sim 10^{50}-10^{51}$ ergs for a SN with kinetic energy $\sim 10^{51}-10^{52}$ ergs, substantially more optically-energetic than a normal Type II SN.  For lower mass or radially-extended disks, the light curve directly traces the shock-generated thermal energy ($\S\ref{sec:fastcool}$; Fig.~\ref{fig:lc1}), plausibly resulting in copious ``intermediate'' width line emission, as observed in SN IIn.  For more massive, compact disks the shock-generated thermal energy must instead diffuse out of the entire stellar envelope ($\S\ref{sec:slowcool}$; Fig.~\ref{fig:lc2}), which may explain the most luminous events (e.g. 2005ap), which show little or no direct evidence for CSM interaction.

Although there is no observational evidence for long-lived disks around massive stars, distinguishing between a truly young stellar object proto-star and an older system with a relic disk may be nontrivial.  Ultimately testing whether bona fide disks exist around evolved stars in sufficient numbers to explain VLSNe will require acquiring unbiased samples of embedded evolved stars and hyper-compact HII regions, distinguishing evolved from youthful systems, and, eventually, pinning down definitive disk or outflow signatures with e.g. atomic \citep{Ercolano&Owen10} and molecular line diagnostics (e.g.~\citealt{Krumholz+07}; \citealt{Carr&Najita08}).  Though challenging, such a feat will be aided by the improved capabilities of present and upcoming observatories such as the EVLA, ALMA, {\it Herschel}, and the {\it James Webb Space Telescope}.

\section*{Acknowledgments}

I thank E.~Quataert, K.~Kratter, N.~Smith, J.~Nordhaus, T.~Thompson, K.~Heng, and R.~Rafikov for carefully reading the text and for providing helpful comments and criticisms.  I also thank J.~Stone, J.~Goodman, A.~Miller, A.~Socrates, R.~Murray-Clay, J.~Converse, D.~Giannios, C.~Wheeler, L.~Hartmann, and R.~Fernandez for helpful conversations.  Support for this work was provided by NASA through Einstein Postdoctoral Fellowship grant number PF9-00065 awarded by the Chandra X-ray Center, which is operated by the Smithsonian Astrophysical Observatory for NASA under contract NAS8-03060.

\begin{appendix}

\section{Calibration of the Ring Model}
\label{appendix:ring}

This Appendix describes how the constants $A$, $B$, and $f$ employed in the ring model in $\S\ref{sec:models}$ are determined.  Most of this material is from Metzger, Piro, $\&$ Quataert (2008), but we repeat it here for convenience.    

The surface density $\Sigma$ of an axisymmetric disk in a Keplerian
potential with constant total angular momentum evolves according to a
diffusion equation (e.g., Frank et al.~2002): \be \frac{\partial
\Sigma}{\partial t} = \frac{3}{r}\frac{\partial}{\partial
r}\left[r^{1/2}\frac{\partial}{\partial r}\left(\nu\Sigma
r^{1/2}\right)\right],
\label{eq:sigma_evo}
\ee where $\nu$ is the kinematic viscosity.  Assuming that $\nu$
depends only on radius as a power law, viz: $\nu =
\nu_{0}(r/R_{d,0})^{n}$, equation (\ref{eq:sigma_evo}) is linear and,
for an initial surface density distribution $\Sigma(r,t=0) =
(M_{d,0}/2\pi R_{d,0})\delta(r-R_{d,0})$ which is narrowly peaked about the
radius $R_{d,0}$, the solution (for $n < 2$) is given by \bea \Sigma(r,t) = \nonumber \eea \be
\frac{M_{d,0}(1-n/2)}{\pi
R_{d,0}^{2}x^{(n+1/4)}\tau}\exp\left[\frac{-(1+x^{2-n})}{\tau}\right]I_{1/|4-2n|}\left[\frac{2x^{1-n/2}}{\tau}\right], \nonumber \\
\label{eq:sigma}
\ee where $M_{d,0}$ is the initial disk mass, $x \equiv r/R_{d,0}$, $\tau
\equiv t[12\nu_{0}(1-n/2)^{2}/R_{d,0}^{2}]$, and $I_{m}$ is a modified
Bessel function of order $m$.  For small argument $y \ll 1$,
$I_{m}(y)$ takes the asymptotic form $I_{m} \simeq
(y/2)^{m}/\Gamma(m+1)$, where $\Gamma$ is the Gamma function; thus,
for late times or small radii such that $\tau \gg 2x^{1-n/2}$,
equation (\ref{eq:sigma}) reduces to \bea \Sigma(r,t)|_{\tau \gg
2x^{1-n/2}} = \nonumber \eea \bea \frac{M_{d,0}}{\pi
R_{d,0}^{2}}\frac{(1-n/2)}{\Gamma[\frac{5-2n}{4-2n}]}\frac{1}{\tau^{\left(\frac{5-2n}{4-2n}\right)}x^{n}}\exp\left[\frac{-(1+x^{2-n})}{\tau}\right]
\label{eq:sigma_asym}
\eea Most of the mass in the disk is located near the radius where the
local mass $M_{d} \propto \Sigma r^{2}$ peaks; using equation
(\ref{eq:sigma_asym}), at late times this radius is found to be
$r_{\rm peak} = R_{d,0}\tau^{1/(2-n)}$.  Hence, equation
(\ref{eq:sigma_asym}) becomes valid near $r_{\rm peak}$ for $\tau \gg
1$.

The constant $A$, which relates the total disk mass at late times from
the exact solution of equation (\ref{eq:sigma_evo}) to the mass
defined by $\pi\Sigma(r_{\rm peak})r_{\rm peak}^{2}$, can
be calculated from equation (\ref{eq:sigma_asym}) to be \be A(\tau \gg
1) \equiv \left.\frac{\int_{0}^{\infty}2\pi \Sigma r
dr}{\pi\Sigma(r_{\rm peak})r_{\rm peak}^{2}}\right|_{\tau \gg 1} =
\frac{2e}{2-n}
\label{eq:a_const}
\ee Similarly, the constant $B$, which relates the total disk angular
momentum at late times from the exact solution to that estimated by
$\pi\Sigma r_{\rm peak}^{2}(GMr_{\rm peak})^{1/2}$, is
given by \be B(\tau \gg 1) \equiv \left.\frac{\int_{0}^{\infty}2\pi
\Sigma r^{3/2} dr}{\pi\Sigma(r_{\rm peak})r_{\rm
peak}^{5/2}}\right|_{\tau \gg 1} =
\frac{2e}{2-n}\Gamma\left[\frac{5-2n}{4-2n}\right]
\label{eq:b_const}
\ee 

From mass continuity, the radial velocity is given by
\be
v_{r} = \frac{-3}{\Sigma r^{1/2}}\frac{\partial}{\partial r}\left[\nu\Sigma r^{1/2}\right] = \frac{-3\nu_{0}}{R_{d,0}}\frac{1}{\Sigma x^{1/2}}\frac{\partial}{\partial x}\left[\Sigma x^{n+1/2}\right],
\label{eq:velocity}
\ee
which, using equation (\ref{eq:sigma_asym}), gives the accretion rate at small radii
\bea \dot{M}_{\rm in} &=& -2\pi\Sigma r v_{r}|_{\tau \gg 2x^{1-n/2}} \nonumber \\ &=& \frac{M_{d,0}}{R_{d,0}^{2}/\nu_{0}}\frac{3(1-n/2)}{\Gamma[(5-2n)/(4-2n)]}\exp[-1/\tau]\tau^{-\left(\frac{5-2n}{4-2n}\right)} \nonumber \\
\label{eq:mdot_analytic}
\eea
Equation (\ref{eq:mdot_analytic}) is easily checked by noting that $\int_{0}^{\infty}\dot{M}_{\rm in}dt = M_{d,0}$, which shows that the entire initial disk eventually accretes onto the central object.  In $\S\ref{sec:ring}$ we introduced the following prescription for evolving the disk mass:
\be
\dot{M}_{d} = \frac{fM_{d}}{t_{\rm visc}},
\label{eq:mdot_f}
\ee where, in terms of the viscosity prescription adopted above,
$t_{\rm visc} = R_{d}^{2}/\nu = t_{\rm visc,0}(R_{d}/R_{d,0})^{2-n}$ and
$t_{\rm visc,0} \equiv R_{d,0}^{2}/\nu_{0}$ is the initial viscous time.
Assuming that the total disk angular momentum remains constant, viz.~$J
\propto M_{d,0}R_{d,0}^{1/2} = (B/A)M_{d}R_{d}^{1/2}$ the solution to equation (\ref{eq:mdot_f}) is given by \be M_{d}(t) = M_{d,0}[1 +
(4-2n)(B/A)^{(4-2n)}f(t/t_{\rm visc,0})]^{-1/(4-2n)}
\label{eq:mdisk}
\ee
In our evolutionary calculations we set $f$ so that the accretion rate from the exact solution to equation (\ref{eq:sigma_evo}) ($\dot{M}_{\rm in}$; eq.~[\ref{eq:mdot_analytic}]) matches the solution to equation (\ref{eq:mdot_f}) at late times (i.e., in the self-similar limit).  This requires
\be
f = 3(1-n/2)
\label{eq:fspecial}
\ee

\subsection{Irradiation-Supported Accretion}

For an irradiation-supported, constant-$\alpha$ disk $\nu \propto H^{2}\Omega \propto r^{15/14}$ (eq.~[\ref{eq:hoverr_irrad}]); thus, $n=15/14$, $f = f_{\rm irr} \simeq 1.4$, $A = A_{\rm irr} \simeq 6.80$, and $B = B_{\rm irr} \simeq 5.85$.  

\subsection{Gravito-turbulent Accretion}

\begin{figure}
\resizebox{\hsize}{!}{\includegraphics[ ]{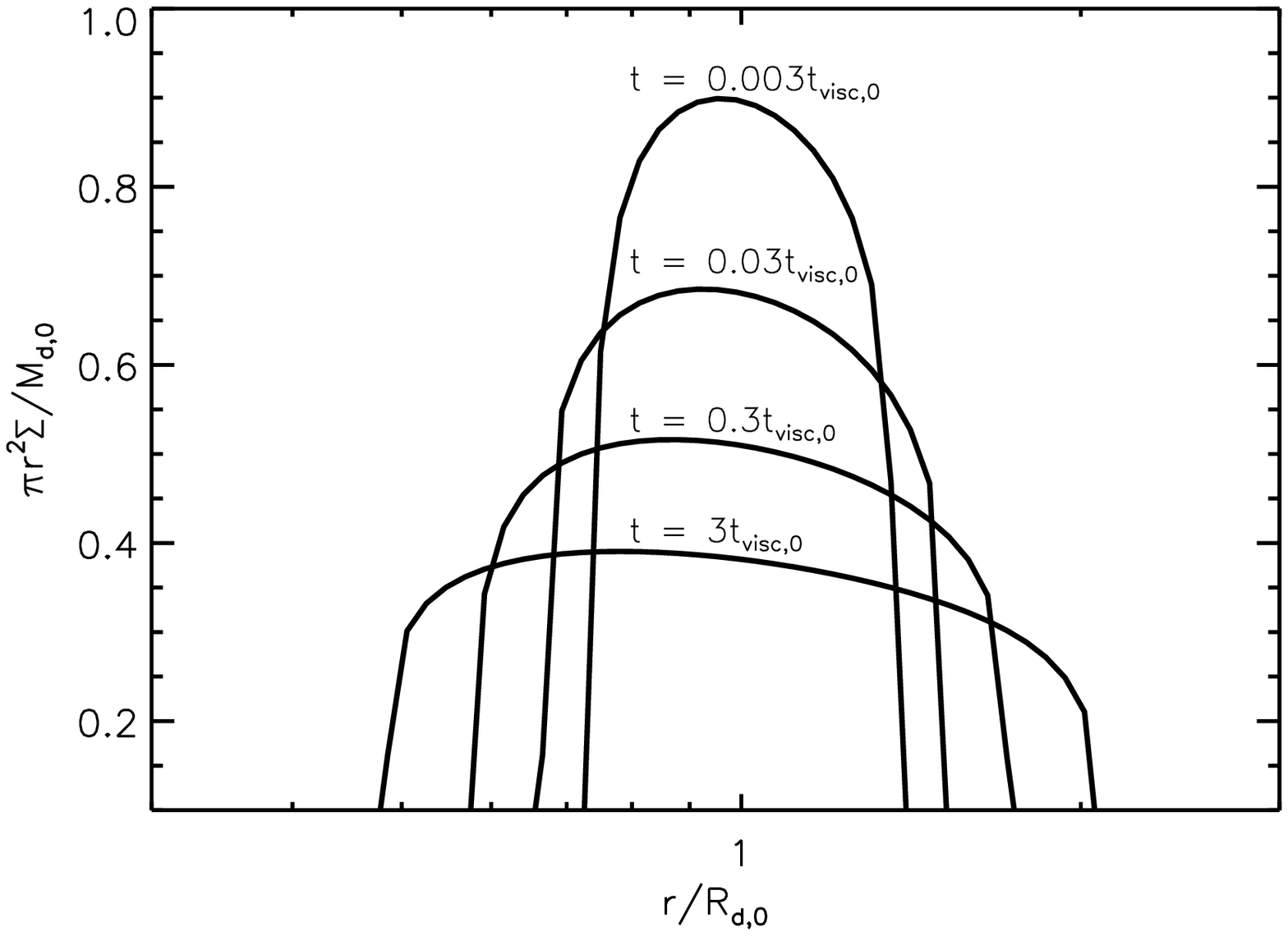}}
\caption{Time evolution of a gravito-turbulent disk with viscosity $\nu \propto \Sigma^{6}r^{15}$, calculated by solving equation (\ref{eq:sigma_evo}) numerically.  The disk is is initially concentrated at the radius $r = R_{d,0}$, and snapshots of the radial profile of the local disk mass $\pi r^{2}\Sigma$ are shown at various times in units of the initial viscous time.  Note that rapidly into the disk's viscous evolution, the disk edges become sharp and mass is spread out approximately equally per decade in radius.}
\label{fig:diskevo}
\end{figure}

For a gravito-turbulent disk $\nu = \alpha_{\rm gi} c_{s}H \propto \Sigma^{6}r^{15}$ ($\S\ref{sec:GIphase}$).  Since $\nu$ in this case depends explicitly on $\Sigma$ as well as $r$, equation (\ref{eq:sigma_evo}) is a no longer linear and the analytic methods employed above do not strictly apply.  In Figure \ref{fig:diskevo} we show $\Sigma(r,t)$, calculated by solving equation (\ref{eq:sigma_evo}) numerically for a disk with viscosity $\nu \propto \Sigma^{6}r^{15}$ and a mass $M_{d,0}$ initially concentrated at the radius $R_{d,0}$.  We show $\Sigma(r)$ at several times, normalized to the initial viscous time at $R_{d,0}$.  Figure \ref{fig:diskevo} shows that as the disk spreads from its initial configuration, a density profile $\Sigma \propto r^{-15/7}$, corresponding to a radially-constant accretion rate $\dot{M} \propto \nu\Sigma$, is rapidly established between the inner and outer edge of the disk.  Note that the disk mass ($\propto \Sigma r^{2} \propto r^{-1/7}$) is concentrated approximately equally per unit decade in radius.  This implies that the constant $A$ (eq.~[\ref{eq:a_const}]) is well approximated as $A_{\rm gi} \equiv M_{d}/\pi R_{d}^{2} \approx {\rm ln}(R_{d}/R_{\rm in})$, where $R_{\rm in} \approx R_{a} \sim 10-30$ AU is the inner edge of the disk (eq.~[\ref{eq:rion}]).  For a typical value $R_{d} \sim 10R_{\rm in}$, we thus estimate that $A = A_{\rm gi} \approx 2$ for gravito-turbulent disks.  

Although the disk evolution is not strictly described by the analytic formalism applied above to the irradiation-supported case, we may still estimate the value of the constant $f_{\rm gi}$ by recognizing that near the peak radius $R_{d}$ the surface density {\it locally} obeys $\Sigma \propto M_{d}/R_{d}^{2}$, implying that (near the peak) we have $\nu \propto \Sigma^{6}r^{15} \propto M_{d}^{6}r^{3} \propto r^{0}$ for constant $J_{d} \propto M_{d}R_{d}^{1/2}$.  For $n=0$ equation (\ref{eq:fspecial}) gives $f = f_{\rm gi} = 3$, the value which we thus employ in our evolutionary calculations in $\S\ref{sec:models}$.  

We note, however, that it is not clear {\it a priori} that the gravito-turbulent disk evolution is indeed dominated by the viscosity at radii $\approx R_{d}$.  The edges of the disk profile in Figure \ref{fig:diskevo} are very sharp because the viscous spreading time $\propto r^{2}/\nu \propto \Sigma^{-6}$ becomes large in regions where $\Sigma$ becomes low.  Low viscosity near the disk edge also causes the disk to spread slower than the prediction $R_{d} \propto t^{1/2}$ of the self-similar model (see Appendix B), which assumes that the disk evolution is dominated by regions of the disk where the majority of mass and angular momentum reside.  This sluggish evolution is, however, probably unphysical.  For one, the assumption of a gravito-turbulent viscosity becomes invalid near the disk edge in regions where the temperature decreases below the floor set by external radiation (see eq.~[\ref{eq:diskmassratio_irr}] and surrounding discussion).  Furthermore, even the assumptions inherent in a height-integrated accretion model become questionable regions where the surface density decreases so sharply that radial gradients exceed vertical gradients.  Thus, although the qualitative features in Figure \ref{fig:diskfig} such as the flat mass profile and sharp disk edges are plausibly physical, the sluggish evolution is likely an artifact of idealized assumptions.  We conclude that a self-similar model using $t_{\rm visc}$ evaluated at $R \sim R_{d}$ may thus remain a reasonable approximation \citep{Pringle91}.

\section{Analytic Self-Similar Solutions}
\label{sec:analytic}

Neglecting details such as stellar mass-loss, the late-time evolution of the disk properties from our calculations in $\S\ref{sec:models}$ asymptote to analytic self-similar power-law solutions.  

From equation (\ref{eq:mdisk}) we observe that at times $t \gg t_{\rm visc,0}$ the disk mass and radius evolve as
\be
M_{d}/M_{d,0} = C_{\rm M}\left(\frac{t}{t_{\rm visc,0}}\right)^{-1/(4-2n)},
\ee
where
\be
C_{\rm M} = \Gamma\left[\frac{5-2n}{4-2n}\right]^{-1}\left((3/4)(4-2n)^{2}\right)^{-1/(4-2n)}
\ee
and
\begin{eqnarray}
R_{d}/R_{d,0} = \left(\frac{M_{d}}{M_{d,0}}\frac{A}{B}\right)^{-2} = C_{\rm R}\left(\frac{t}{t_{\rm visc,0}}\right)^{1/(2-n)},
\label{eq:rratio}
\end{eqnarray}
where 
\be
C_{\rm R} = \Gamma\left[\frac{5-2n}{4-2n}\right]^{4}\left((3/4)(4-2n)^{2}\right)^{1/(2-n)}
\ee
and equation (\ref{eq:rratio}) follows from angular momentum conservation.

For an irradiation-supported disk $n =15/14$, $M_{d} \propto t^{-7/13}$ and $R_{d} \propto t^{14/13}$.  Although gravito-turbulent disks obey a non-linear diffusion equation, we find $n = 0$ using disk properties evaluated locally at radii $\sim R_{d}$ (see Appendix A).  In this case we thus have $M_{d} \propto t^{-1/4}$ and $R_{d} \propto t^{1/2}$.  As described in $\S\ref{sec:models}$, all of the other disk properties at $R = R_{d}$ follow from $M_{d}(t)$ and $R_{d}(t)$.  The full set of self-similar solutions for the gravito-turbulent and irradiation-supported $\alpha$-disk are given in equations [\ref{eq:MdGIss}]-[\ref{eq:omegatcoolGIss}] and [\ref{eq:Mdirrss}]-[\ref{eq:Qirrss}], respectively. 

\end{appendix}




\label{lastpage}

\end{document}